\documentclass[12pt]{article}
\usepackage{enumerate}
\usepackage{amsfonts}
\usepackage{amsmath}
\usepackage{amssymb}
\usepackage{amsthm}
\usepackage{latexsym}
\usepackage{color}
\usepackage{graphicx}
\usepackage{wrapfig}
\usepackage{caption}
\usepackage{float}

\def\D{\mathcal{D}}
\def\H{\mathcal{H}}

\def\M{\mathfrak{M}}

\def\S{\mathfrak{S}}

\def\C{\mathfrak{C}}
\def\T{\mathfrak{T}}

\def\N{\mathbb{N}}

\newcommand{\supp}{\mathrm{supp}}
\newcommand{\rank}{\mathrm{rank}}

\newcommand{\Tr}{\mathrm{Tr}}

\newcommand{\shs}{\hspace{1pt}}
\newcounter{defin}  \newcounter{lemma}  \newcounter{theorem}
\newcounter{proposition} \newcounter{corol}  \newcounter{remark} \newcounter{example}
\newenvironment{lemma}{\par\refstepcounter{lemma}     \textbf{Lemma \thelemma.} }{\rm\par}
\newenvironment{theorem}{\par\refstepcounter{theorem}     \textbf{Theorem \thetheorem.}\ }{\rm\par}
\newenvironment{proposition}{\par\refstepcounter{proposition}     \textbf{Proposition \theproposition.}\ }{\rm\par}
\newenvironment{corollary}{\par\refstepcounter{corol}     \textbf{Corollary \thecorol.} }{\rm\par}

\newenvironment{remark}{\par\refstepcounter{remark}     \textbf{Remark \theremark.}}{\rm\par}
\newenvironment{example}{\par\refstepcounter{example}     \textbf{Example \theexample.}}{\rm\par}

\begin{document}


\title{Optimal Hamiltonian for a quantum state with finite entropy}


\author{M.E.~Shirokov\footnote{email:msh@mi.ras.ru}\\
Steklov Mathematical Institute, Moscow, Russia}
\date{}
\maketitle
\begin{abstract}
We consider the following task: how for a given quantum state $\rho$  to find a grounded Hamiltonian $H$ satisfying the condition $\mathrm{Tr} H\rho\leq E_0<+\infty$ in such a way that
the von Neumann entropy of the Gibbs state $\gamma_H(E)$ corresponding to a given energy $E>0$ be as small as possible.

We show that for any mixed state $\rho$ with finite entropy and any $E>0$ there exists a  solution $H(\rho,E_0,E)$ of the above problem (unique in the non-degenerate case) which we call  optimal Hamiltonian
for the state $\rho$. Explicit expressions for $H(\rho,E_0,E)$, $\gamma_{H(\rho,E_0,E)}(E)$ and $S(\gamma_{H(\rho,E_0,E)}(E))$  are obtained. Analytical properties of the function $E\mapsto S(\gamma_{H(\rho,E_0,E)}(E))$ are explored. Several examples are considered.

We also consider a modification of the above task in which arbitrary Hamiltonians (not necessarily grounded) are considered.

The basic application motivating this research is described. As examples, new semicontinuity bounds for the von Neumann entropy and for the entanglement of formation are obtained and briefly discussed (with the intention to give a detailed analysis in a separate article).
\end{abstract} 

\tableofcontents

\section{Introduction}

Linear (affine) constraints on a quantum state $\rho$ induced by the inequalities
$$
\Tr H_1\rho\leq E_1,...,\Tr H_m\rho\leq E_m,\quad E_1,...,E_m\in \mathbb{R},
$$ where $H_1$,...,$H_m$ are Hermitian (in particular, positive) operators
are widely used in quantum information theory \cite{H-SCI,N&Ch,Wilde}. They are noncommutative analogs of the constraints on a probability distribution $\{p_i\}_{i\in I}$
given by the inequalities
$$
\sum_{i\in I}a^1_ip_i\leq E_1,...,\sum_{i\in I}a^m_ip_i\leq E_m,\quad E_1,...,E_m\in \mathbb{R},
$$ where $\{a^1_i\}_{i\in I}$,...,$\{a^m_i\}_{i\in I}$ are sets of real (in particular, nonnegative) numbers \cite[Section 3.4]{Moser}.

Typically, the constraint operators $H_1$,...,$H_m$ are observables of physical quantities (energy, position, momentum, etc.) \cite[Ch.12]{H-SCI},\cite{W}. At the same time,
in different tasks of quantum information theory  artificially constructed  constraint operators $H_1$,...,$H_m$ with specific
properties may be useful (see, f.i., the proof of the converse part of Theorem 7 in \cite{Lami-new}). The linear constrains are applied in the quantum communication theory \cite{H-SCI}, the entanglement theory \cite{4H}, semidefinite programming \cite[Section 2.4]{Wilde-new}, etc.

Starting from  Winter's work \cite{W-CB}, the linear constraints are essentially used in quantitative
continuity analysis of entropic and of information characteristics of an infinite-dimensional $n$-partite quantum system $A_1...A_n$. Within this application
the constraint operators $H_1$,...,$H_m$ are assumed to be positive and treated as Hamiltonians (energy observables) of the subsystems $A_1$,..,$A_m$, $m\leq n$ (see \cite{QC} and the references therein).

In this direction, the notion of\emph{ lower semicontinuity bound }(LSB) naturally appeared. If $f(\rho)$ is a function on some subset $\S_0$ of the set $\S(\H)$ of quantum states taking values in $(-\infty,+\infty]$ then a LSB for this function at a state $\rho$ with finite $f(\rho)$ is the inequality
\begin{equation}\label{SCB-g}
f(\rho)-f(\sigma)\leq \mathrm{LSB}_f(\varepsilon|\shs \rho)
\end{equation}
valid for any $\sigma$ in $\S_0\cap U_{\varepsilon}(\rho)$, where $U_{\varepsilon}(\rho)$ is the $\varepsilon$-vicinity of the state $\rho$ w.r.t. some metric on $\S(\H)$ and
it is assumed that the function $\mathrm{LSB}_f(\varepsilon|\shs \rho)$ may depend on $\rho$ but cannot depend on $\sigma$. The function $f(\rho)$ is
lower semicontinuous at a state $\rho$ if and only if (\ref{SCB-g}) holds with  the function $\mathrm{LSB}_f(\varepsilon|\shs \rho)$ tending to zero as $\,\varepsilon\to0$.
In this case the LSB (\ref{SCB-g}) is called \emph{faithful}.

Despite the fact that many characteristics of quantum systems are lower semicontinuous functions either on $\S(\H)$ or on some subset of $\S(\H)$, the problem
of finding LSB for these characteristics in the case $\dim\H=+\infty$ is not trivial. A short list of characteristics of a quantum $n$-partite system $A_1...A_n$ for which LSB are constructed up to now can be found in Section 6. A common feature of all these LSB is the appearance of the main term
\begin{equation}\label{mt}
C \varepsilon F_H\!\left(\frac{E}{\varepsilon}\right),\quad C>0,
\end{equation}
in the expression for $\mathrm{LSB}_f(\varepsilon|\shs \rho)$. Here, $H$ is a positive operator on $\H_{A_1}$ with particular properties\footnote{discrete spectrum of finite multiplicity with
sufficiently large increasing rate and the minimal eigenvalue equal to zero (see Section 2)} (it is treated as  Hamiltonian of a quantum system described by the Hilbert space $\H_{A_1}$), $E=\Tr H\rho_{A_1}$ (assumed to be finite) and $F_H(E)$ is the von Neumann entropy of the Gibbs state $\gamma_H(E)$ corresponding
to the "energy" $E$. For any state $\rho$ with finite  $\,S(\rho_{A_1})\,$ a positive operator $H$ such that $\,\Tr H\rho_{A_1}<+\infty\,$  can be found by uncountable number of ways. So, the question arises \emph{how for given $\rho$ and $\,\varepsilon>0$ to find
a positive operator $H$ with finite $\Tr H\rho_{A_1}$ in such a way that the term (\ref{mt})
be as small as possible.} It was this question that motivated the research presented below.\smallskip

Keeping in mind potential applications beyond the problem stated before, we formulate the main task of this article in the following form:
\emph{how for a given quantum state $\rho$  to find a grounded Hamiltonian $H$ with $\Tr H\rho\leq E_0$ in such a way that
the value of $\,F_H(E)\doteq S(\gamma_H(E))\,$ for a given  $E>0$ be as small as possible.}

We show that for any mixed state $\rho$ with finite entropy and any $E>0$  there are two possibilities:
\begin{enumerate}[A)]
  \item any grounded Hamiltonian $H$ with the domain $\,\supp\rho\,$ such that $\Tr H\rho\leq E_0$  gives a solution of the above problem (this holds if $\rank\rho<+\infty$ and $E$ is large enough);
  \item there exists a unique solution $H(\rho,E_0,E)$ of the above problem.
\end{enumerate}
We describe  explicit expressions for $H(\rho,E_0,E)$, $\gamma_{H(\rho,E_0,E)}(E)$ and $S(\gamma_{H(\rho,E_0,E)}(E))$ in Section 3, analyse properties of the function $E\mapsto S(\gamma_{H(\rho,E_0,E)}(E))$  in Section 4 and consider concrete examples  in Section 5. 

In Section 6, we consider a modification of the above optimization problem  in which arbitrary Hamiltonians (not necessarily grounded) are considered.

In Section 7, a brief overview of the basic application motivating this research is given and new semicontinuity bounds for the von Neumann entropy and for the entanglement
of formation are obtained and briefly discussed. 

The Appendix contains basic lemmas and  the axillary results used in the main part of the article.

\section{Preliminaries}

Let $\mathcal{H}$ be a separable infinite-dimensional Hilbert space,
$\mathfrak{B}(\mathcal{H})$ the algebra of all bounded operators on $\mathcal{H}$ with the operator norm $\|\cdot\|$ and $\mathfrak{T}( \mathcal{H})$ the
Banach space of all trace-class
operators on $\mathcal{H}$  with the trace norm $\|\!\cdot\!\|_1$. Let
$\mathfrak{S}(\mathcal{H})$ be  the set of quantum states (positive operators
in $\mathfrak{T}(\mathcal{H})$ with unit trace) \cite{H-SCI,N&Ch,Wilde}.


The \emph{support} $\mathrm{supp}\rho$ of an operator $\rho$ in  $\T_{+}(\H)$ is the closed subspace spanned by the eigenvectors of $\rho$ corresponding to its positive eigenvalues.  The dimension of $\mathrm{supp}\rho$ is called the \emph{rank} of $\rho$ and is denoted by $\rank\rho$.\smallskip

We will use the Mirsky inequality
\begin{equation}\label{Mirsky-ineq+}
  \sum_{i=1}^{+\infty}|\lambda^{\rho}_i-\lambda^{\sigma}_i|\leq \|\rho-\sigma\|_1
\end{equation}
valid for any positive operators $\rho$ and $\sigma$ in $\T(\H)$, where  $\{\lambda^{\rho}_i\}_{i=1}^{+\infty}$
and $\{\lambda^{\sigma}_i\}_{i=1}^{+\infty}$ are the sequences
of eigenvalues of $\rho$ and $\sigma$ arranged in the non-increasing order (taking the multiplicity into account) \cite{Mirsky,Mirsky-rr}.\smallskip

The \emph{von Neumann entropy} of a quantum state
$\rho \in \mathfrak{S}(\H)$ is  defined by the formula
$S(\rho)=\operatorname{Tr}\eta(\rho)$, where  $\eta(x)=-x\ln x$ if $x>0$
and $\eta(0)=0$. It is a concave lower semicontinuous function on the set~$\mathfrak{S}(\H)$ taking values in~$[0,+\infty]$ \cite{H-SCI,L-2,W}.\smallskip

In the article, we will use the class $\mathfrak{H}(\H)$ of positive (semi-definite)  operators on a Hilbert space $\mathcal{H}$
with discrete spectrum of finite multiplicity (called \emph{$\mathfrak{H}$-operators} in what follows). There are two types of $\mathfrak{H}$-operators:
\begin{enumerate}[A)]
  \item positive (semi-definite)  operators $H$ with finite-dimensional domain $\D(H)$, which can be treated as finite-dimensional positive operators on  $\D(H)$ with
  the spectral representation
  $$
  H=\sum_{k=1}^{n} h_k |\tau_k\rangle\langle\tau_k|,\quad n=\dim \D(H),
  $$
  $\mathcal{T}\doteq\left\{\tau_k\right\}_{k=1}^{n}$ is a basis of eigenvectors of $H$ in $\D(\H)$ and $\left\{h_k\right\}_{k=1}^{n}$ is the corresponding set of its eigenvalues (which is assumed to be nondecreasing);
  \item positive (semi-definite)  operators $H$ with infinite-dimensional domain $\D(H)$ and
  the spectral representation
  \begin{equation}\label{H-form}
H=\sum_{k=1}^{+\infty} h_k |\tau_k\rangle\langle\tau_k|,
\end{equation}
where $\mathcal{T}\doteq\left\{\tau_k\right\}_{k=1}^{+\infty}$ is the orthonormal
system of eigenvectors of $H$ corresponding to the \emph{nondecreasing} unbounded sequence $\left\{h_k\right\}_{k=1}^{+\infty}$ of its eigenvalues; in this case
$\D(H)=\{ \varphi\in\H_\mathcal{T}\,| \sum_{k=1}^{+\infty} h^2_k |\langle\tau_k|\varphi\rangle|^2<+\infty\}$, where $\H_\mathcal{T}$ of the linear span of $\mathcal{T}$.
\end{enumerate}

For any $\mathfrak{H}$-operator $H$ and positive operator $\rho\in\T(\H)$ we will define the quantity $\Tr H\rho$ by the rule
\begin{equation*}
\Tr H\rho=
\left\{\begin{array}{ll}
        \Tr H\rho\; &\textrm{if}\;\;  \dim \mathcal{D}(H)<+\infty\quad \textrm{and}\quad  \supp\rho\subseteq \mathcal{D}(H)\\
        \sup_m \Tr P_m H\rho\; &\textrm{if}\;\; \dim \mathcal{D}(H)=+\infty \quad \textrm{and}\quad\supp\rho\subseteq {\rm cl}(\mathcal{D}(H))\\
        +\infty\;\;&\textrm{otherwise}
        \end{array}\right.,
\end{equation*}
where $P_m=\sum_{k=1}^{m}|\tau_k\rangle\langle\tau_k|$ is the spectral projector of $H$ corresponding to the  $m$ maximal eigenvalues of $H$ and ${\rm cl}(\mathcal{D}(H))$ is the closure of $\mathcal{D}(H)$.

To unify description of $\mathfrak{H}$-operators of types $A$ and $B$ we will assume
that\break $\mathfrak{H}$-operators of type $A$ have representation (\ref{H-form}) in which all the "eigenvalues" $h_k$ with $k$ greater than $\,n=\dim \D(H)\,$ are equal to $+\infty$.
Such assumption seems natural, since the $\mathfrak{H}$-operators of type $A$ can be treated as limit points of increasing sequences of operators of type $B$.

\textbf{Note:} \emph{To make the expression $\Tr e^{-\beta H}$ (and  similar expressions) be valid in this case with $\beta=0$  we will assume that $\,0\cdot h_k=+\infty$ for any $h_k=+\infty$.}

Within this convention we have
\begin{equation}\label{tpt}
\Tr\, e^{-\beta H}=\sum_{k=1}^{+\infty}e^{-\beta h_k}\quad \textrm{and} \quad  \Tr H \rho=\sum_i p_i\|\sqrt{H}\varphi_i\|^2\leq+\infty
\end{equation}
for any operator $\rho$ in $\T_+(\H)$ with the spectral decomposition $\rho=\sum_i p_i|\varphi_i\rangle\langle\varphi_i|$ provided that
all the eigenvectors $\varphi_i$ lie in $\D(\sqrt{H})=\{ \varphi\in\H_\mathcal{T}\,| \sum_{k=1}^{+\infty} h_k |\langle\tau_k|\varphi\rangle|^2<+\infty\}$. If at least one eigenvector of $\rho$ corresponding to a nonzero eigenvalue does not belong to the set $\D(\sqrt{H})$ then $\Tr H \rho=+\infty$.


We may treat an $\mathfrak{H}$-operator $H$ on a Hilbert space $\H$ as the Hamiltonian (energy observable) of a quantum system
described by the  space $\H$. In this case $\,\Tr H \rho\,$ is the mean energy of a quantum state $\rho$.

For any $\mathfrak{H}$-operator $H$ the set
$$
\C_{H,E}=\left\{\rho\in\S(\H)\,|\,\Tr H\rho\leq E\right\}
$$
is convex and closed (since the function $\rho\mapsto\Tr H\rho$  is affine and lower semicontinuous). It is nonempty if $E$ is greater than the infimum of the spectrum of $H$.

It is easy to show  (see, f.i., Proposition 1 in \cite{EC}) that
\begin{equation}\label{F-def}
F_{H}(E)\doteq\sup_{\rho\in\C_{H,E}}S(\rho)<+\infty
\end{equation}
for any $E>0$ if and only if the $\mathfrak{H}$-operator $H$ satisfies  the condition
\begin{equation}\label{H-cond-w}
  \Tr\, e^{-\beta H}<+\infty\quad\textrm{for some }\;\,\beta>0.
\end{equation}
Proposition 1 in \cite{EC} also implies that the more strong  condition
\begin{equation}\label{H-cond}
  \Tr\, e^{-\beta H}<+\infty\quad\textrm{for all }\;\,\beta>0
\end{equation}
is equivalent to the following asymptotic property
\begin{equation}\label{H-cond-a}
  F_{H}(E)=o\shs(E)\quad\textrm{as}\quad E\rightarrow+\infty.
\end{equation}

For $\mathfrak{H}$-operators of type $A$ condition (\ref{H-cond-w}) and (\ref{H-cond}) hold trivially.\smallskip

If there is a state at which the supremum in (\ref{F-def}) is attained then this state is always unique (by the strict concavity of the entropy) and called the \emph{Gibbs state} of the
operator $H$ corresponding to the "energy" $E$ \cite{W}. We will denote it by $\gamma_H(E)$.

If $H$ is an $\mathfrak{H}$-operator of type $A$ ($\dim\D(H)=n<+\infty$, $h_k=+\infty$ for all $k>n$) then the Gibbs state $\gamma_H(E)$ always exists (by simple finite-dimensional arguments). If $E\leq h_*(H)\doteq\frac{1}{n}\sum_{k=1}^nh_k$ then
\begin{equation}\label{Gibbs}
\gamma_H(E)=\frac{e^{-\beta_H(E) H}}{\Tr e^{-\beta_H(E)H}},
\end{equation}
where the parameter $\beta_H(E)\geq0$ is determined by the equation\footnote{See Appendix A-2.}
\begin{equation}\label{imp-in}
\Tr H e^{-\beta H}=E\Tr e^{-\beta H}.
\end{equation}
If $\,E\geq h_*(H)\,$  then $$\gamma_H(E)=\frac{1}{n}\sum_{k=1}^n|\tau_k\rangle\langle\tau_k|.$$
\smallskip

If $\,H$ is an $\mathfrak{H}$-operator of type $B$ ($\dim\D(H)=+\infty$, $h_k<+\infty\,$ for all $\,k$) then\break Proposition 1 in \cite{EC} shows that the Gibbs state $\gamma_H(E)$ exists if and only if\break $E\in[h_1,h_*(H)]$, where
$$
h_*(H)\doteq\left\{\begin{array}{ll}
        \frac{\Tr H e^{-g(H)H}}{\Tr e^{-g(H)H}}\; &\textrm{if}\;\;  \Tr e^{-g(H)H}<+\infty\medskip\\
        +\infty\; &\textrm{if}\;\;  \Tr e^{-g(H)H}=+\infty
        \end{array}\right.\!,\quad g(H)=\inf\left\{\beta\in\mathbb{R}_+|\,\Tr\, e^{-\beta H}<+\infty\right\}\!,
$$
If $E\in(h_1,h_*(H)]$ then the Gibbs state $\gamma_H(E)$ is given by formula (\ref{Gibbs}), where the parameter $\beta_H(E)>0$ is determined by equation
(\ref{imp-in}).\smallskip

In this article, we will use the function
\begin{equation}\label{Z}
  Z_H(E) \doteq \Tr e^{-\beta_H(E)H}=\sum_{k=1}^{+\infty}e^{-\beta_H(E)h_k}.
\end{equation}
where the parameter $\beta_H(E)\geq0$ is determined by the equation (\ref{imp-in}). It is easy to see that $Z_H(E)$ is an increasing function on  $[h_1,+\infty)$ (because $\beta_H(E)$ is a decreasing function).\smallskip

We will use the following observation, which can be established using the  proof of Proposition 1 in \cite{EC}.\smallskip
\begin{lemma}\label{Gibbs-r} \emph{If equality (\ref{imp-in}) holds for some $\beta>0$ then the state proportional to the operator $e^{-\beta H}$
is the Gibbs state $\gamma_H(E)$. If $\,H$ is an $\mathfrak{H}$-operator of type $B$ ($\dim\D(H)=+\infty$) then the existence of the Gibbs state $\gamma_H(E)$ implies that this state has the form (\ref{Gibbs}) and that the equality (\ref{imp-in}) holds with $\beta=\beta_H(E)$.}
\end{lemma}\medskip

The function $F_{H}(E)$ defined in (\ref{F-def}) is strictly concave on $[h_1,h_*(H)]$ and linear on $[h_*(H),+\infty)$  \cite[Proposition 1]{EC}.
We will often assume that
\begin{equation}\label{star}
  h_1\doteq\inf\limits_{\|\varphi\|=1}\langle\varphi\vert H\vert\varphi\rangle=0.
\end{equation}

We will essentially use the following representation for the function $F_{H}(E)$.\smallskip

\begin{lemma}\label{V} \emph{Let $H$ be an arbitrary $\mathfrak{H}$-operator with representation (\ref{H-form}). Then}
$$
F_H(E)=\inf_{\beta\geq 0}\left( E\beta +\ln\Tr e^{-\beta H}\right)=\inf_{\beta\geq 0}\left( E\beta+\ln\sum_{k=1}^{+\infty}e^{-\beta h_k}\right).
$$
\end{lemma}
If $H$ is an $\mathfrak{H}$-operator of type $B$ then the claim of Lemma \ref{V} follows from Proposition 1 in \cite{EC}. For $\mathfrak{H}$-operators of type $A$
the proof can be essentially simplified. It is placed in Appendix A-2.\smallskip


\textbf{Note:} Throughout the article  we use the term  "$n$-tuple of numbers" in both cases $\,n\in\N\,$ and $\,n=+\infty\,$ simultaneously. In the second case we assume that it is a countable ordered subset of $\mathbb{R}$ (a sequence).

\section{The main results}

To formulate our main results we need the maps $\mu_{\rm c}$, $\langle\cdot\rangle_{\rm c}$, $\mu_{\rm q}$ and $\langle\cdot\rangle_{\rm q}$ described below.  \smallskip

Let $n$ be either a natural number greater than $1$ or $+\infty$. The maps $\mu_{\rm c}$ and $\langle\cdot\rangle_{\rm c}$ transform a non-increasing $n$-tuple\footnote{According to the notation introduced in Section 2 we assume that $n$-tuple with $n=+\infty$ is a countable ordered subset of $\mathbb{R}$ (a sequence).} $\,\bar{x}=\{x\}_{i=1}^{n}\,$ of positive numbers with finite $\,\|\bar{x}\|_1\doteq\sum_{i=1}^{n}x_i\,$ into, respectively,
a natural number and a probability distribution with $n$ outcomes according to the rules
\begin{equation}\label{m-c}
\mu_{\rm c}(\bar{x})= \left\{\begin{array}{ll}
        1 &\;\textrm{if}\;\;  \|\bar{x}\|_1\leq 1 \\\\
        \max\{k\leq n\,|\,\sum_{i=1}^{k}x_i-kx_k\leq\|\bar{x}\|_1-1\} &\;\textrm{if}\;\;  \|\bar{x}\|_1>1
        \end{array}\right.\!,
\end{equation}
\begin{equation}\label{psi-c}
\langle\bar{x}\rangle_{\rm c}= \left\{\begin{array}{ll}
        \left(\left(1-\sum_{i=2}^{n}x_i\right),x_2,x_3,...\right)&\;\textrm{if}\;\;  \|\bar{x}\|_1\leq 1 \\\\
        (\underbrace{c,...,c}_{m\textrm{ items}}, x_{m+1},x_{m+2},...),\quad m=\mu_{\rm c}(\bar{x}),&\;\textrm{if}\;\;  \|\bar{x}\|_1>1
        \end{array}\right.\!,
\end{equation}
where $\,c=\frac{1-\sum_{i=m+1}^{n}x_i}{m}$. It is easy to see that the $n$-tuple of numbers $\,c_k\doteq\sum_{i=1}^{k}x_i-kx_k\,$ is
nondecreasing, $c_1=0$ and $c_k$ tends to $\|\bar{x}\|_1$ as $k\to+\infty$ in the case $n=+\infty$. So, the map $\mu_{\rm c}$
is well defined.\smallskip

Roughly speaking, the action of the map $\langle\cdot\rangle_{\rm c}$ is the construction of a probability
distribution from an $n$-tuple $\,\bar{x}=\{x\}_{i=1}^{n}\,$ by
\begin{itemize}
  \item increasing of the first entry of $\,\bar{x}\,$ in the case $\,\|\bar{x}\|_1<1$,
  \item leaving  $\,\bar{x}\,$ as it is in the case $\,\|\bar{x}\|_1=1$,
  \item the uniform truncation of the maximal entries of $\,\bar{x}\,$ (in such a way that
the entries of $\,\langle\bar{x}\rangle_{\rm c}\,$ are arranged in the non-increasing order) in the case $\,\|\bar{x}\|_1>1$.\footnote{It is easy to see that $\,c\geq x_{m+1}$.}
\end{itemize}

The maps $\mu_{\rm q}$ and $\langle\cdot\rangle_{\rm q}$  (the quantum versions of $\mu_{\rm c}$ and $\langle\cdot\rangle_{\rm c}$) transform a positive (nonzero) operator $\sigma$ in $\T(\H)$ with the spectral representation $\,\sigma=\sum\limits_{i=1}^{n} q_{i}|\psi_i\rangle\langle \psi_i|$, ($q_i>0$, $1<n\leq+\infty$), into, respectively, a natural number and a quantum state by the rules
\begin{equation}\label{psi-q}
\mu_{\rm q}(\sigma)= \mu_{\rm c}(\bar{q}),\quad \langle\sigma\rangle_{\rm q}=\sum\limits_{i=1}^{n} \langle\bar{q}\rangle_{\rm c}^{i}|\psi_i\rangle\langle \psi_i|,\quad \bar{q}=(q_1,q_2,...),
\end{equation}
where $\langle \bar{q}\rangle_{\rm c}^{i}$ is the $i$-th entry of $\langle \bar{q}\rangle_{\rm c}$.

It is not hard to show that the map $\langle \cdot\rangle_{\rm c}$  (resp. the map $\langle \cdot\rangle_{\rm q}$)  is \emph{continuous} w.r.t. the $\ell_1$-norm  (resp. the trace norm $\|\cdot\|_1$).\smallskip

For a given state $\rho\in\S(\H)$ let $\M(\rho,E_0)$ be the set of all $\mathfrak{H}$-operators $H$ on $\H$ (defined in Section 2) satisfying  conditions (\ref{H-cond-w}) and (\ref{star}) such that $\Tr H\rho\leq E_{0}$, $E_0>0$. For given $E>0$ let $\M_E(\rho,E_0)$ be the subset of $\M(\rho,E_0)$ consisting of $\mathfrak{H}$-operators $H$ for each of which there exists the Gibbs state $\gamma_{H}(E)$ corresponding to the "energy" $E$ (see Section 2).\smallskip

\begin{theorem}\label{main}\emph{Let $\H$ be a separable infinite-dimensional Hilbert space and $\rho$ be a mixed state in $\S(\H)$ with the spectral representation $\rho=\sum\limits_{i=1}^{n} p_{i}|\varphi_i\rangle\langle \varphi_i|$ (where $n\leq +\infty$, $p_{i+1}\leq p_i$ and $p_i>0$ for all $\,i$) such that $S(\rho)<+\infty$. Let $\,E_0,E>0$ be arbitrary and $\theta=E/E_0$. Write $A$ and $B$ for the following cases:
\begin{itemize}
 \item $\,n<+\infty\,$ and  $\,E\geq E_0/(p_nn)\;$ (case $A$);
 \item either $\,n=+\infty\,$ or  $\,E<E_0/(p_nn)\;$  (case $B$).
\end{itemize}
Let
\begin{eqnarray*}
  d_k&=&\sum_{i=k+1}^{n}p_i=1-\sum_{i=1}^{k}p_i,\quad k\in\N\cap[1,n),\quad d_n=0\;\;\textit{if}\;\; n<+\infty, \\
  s_k&=&\sum_{i=k+1}^{n}\eta(p_i)=S(\rho)-\sum_{i=1}^{k}\eta(p_i),\quad k\in\N\cap[1,n),\quad s_n=0\;\;\textit{if}\;\; n<+\infty,
\end{eqnarray*}
where $\eta(x)=-x\ln x$.}\smallskip

\emph{There exists an operator $H(\rho,E_0,E)$ in $\M_E(\rho,E_0)$ such that
\begin{equation}\label{main+}
S(\gamma_{H(\rho,E_0,E)}(E))=\inf_{H\in\M_E(\rho,E_0)}S(\gamma_{H}(E))=\inf_{H\in\M(\rho,E_0)}\sup_{\Tr H\sigma\leq E}S(\sigma),
\end{equation}
where the supremum is over all states $\sigma$ in $\S(\H)$ such that $\Tr H\sigma\leq E$.}
\smallskip

\emph{In case $A$ the role of $\,H(\rho,E_0,E)$ can be played by any operator in $\M(\rho,E_0)$ such that $\,\D(H)=\supp\rho\,$, for example, by
the operator $H_\rho$ equal to zero on $\,\D(H_\rho)=\supp\rho\,$ or by the operator\footnote{The sense of this  choice of $H(\rho,E_0,E)$  in case $A$ is explained in Remark \ref{main-r+} below.}
\begin{eqnarray}\label{r-c}
\sum\limits_{i=1}^{+\infty} h^A_{i}|\varphi_i\rangle\langle \varphi_i|,\qquad  h^A_i &=& \left\{\begin{array}{ll}
        0&\textrm{if}\;\;  1\leq i\leq \tilde{m} \\
        \frac{E_0}{p_n(n-\tilde{m})}& \textrm{if}\;\; \tilde{m}<i\leq n \\
        +\infty & \textrm{if}\;\; i> n
        \end{array}\right.\!\!,
\end{eqnarray}
where $\,\tilde{m}=n-\max\{k\in\N\cap[1,n-1]\,|\,p_{n-k+1}=p_n\}$.}\footnote{We use the definition described after (\ref{H-form}) which in this case implies that $\D(H(\rho,E_0,E))=\supp\rho$. We assume that $\{\varphi_i\}_{i=1}^{+\infty}$ is any extension of the system $\{\varphi_i\}_{i=1}^{n}$ to a basis in $\H$.}

\emph{In case $B$  the operator $H(\rho,E_0,E)$ is unique (up to natural isomorphisms), defined on the domain of $\,\ln\rho\,$ and can be represented in the following equivalent forms
\begin{itemize}
  \item \begin{equation}\label{HB-exp-1}
H(\rho,E_0,E)=\frac{E_0(I_{\H}\ln \|\langle\theta\rho\rangle_{\rm q}\| -\ln \langle\theta\rho\rangle_{\rm q})}{\ln\|\langle\theta\rho\rangle_{\rm q}\|-\Tr\rho\ln\langle\theta\rho\rangle_{\rm q}};
\end{equation}
  \item
\begin{equation}\label{HB-exp-2}
 H(\rho,E_0,E)=\frac{P_m}{\beta_m}\left(-\ln\rho+c_mI_\H\right),\quad m=\mu_{\rm q}(\theta\rho);
\end{equation}
  \item \begin{equation}\label{HB-exp-3}
H(\rho,E_0,E)=\sum\limits_{i=1}^{+\infty} h^B_{i}|\varphi_i\rangle\langle \varphi_i|,
\end{equation}
\begin{eqnarray*}
\textit{where}\qquad    h^B_i &=& \left\{\begin{array}{ll}
        0&\textrm{if}\;\;  1\leq i\leq m \\
        \frac{1}{\beta_m}\left(\ln\frac{1-\theta d_m}{\theta m}-\ln p_i\right)& \textrm{if}\;\; m< i\leq n \\
        +\infty & \textrm{if}\;\; n<+\infty\;\;\textit{and}\;\; i> n
        \end{array}\right.\!.
\end{eqnarray*}
\end{itemize}
Here $\,\mu_{\rm q}$ and $\,\langle\cdot\rangle_{\rm q}$ are the maps defined in (\ref{psi-q}), $\|\langle\theta\rho\rangle_{\rm q}\|$ is the operator norm of $\,\langle\theta\rho\rangle_{\rm q}$,
$\,\beta_m=\frac{1}{E_0}(s_m+d_m\ln\frac{1-\theta d_m}{\theta m})$,  $\,c_m=\ln \frac{1-\theta d_m}{\theta m}\,$ and $\;P_m=\sum\limits_{i=m+1}^{n}|\varphi_i\rangle\langle \varphi_i|$.}

\emph{In both cases $A$ and $B$ the state $\,\gamma_{H(\rho,E_0,E)}(E)$ is uniquely defined,\footnote{The equality  $\gamma_{H(\rho,E_0,E)}(E)=\langle\theta\rho\rangle_{\rm q}$ is the most interesting and surprising claim of Theorem 1, it is proved  by a very non-direct way
via  the  coincidence of $S(\gamma_{H(\rho,E_0,E)}(E))$ with $S\!\left(\langle\theta\rho\rangle_{\rm q}\right)$ established
by the solution of the corresponding optimization problem (see the below proof of Theorem \ref{main}). \emph{It would be interesting} to
find a direct way to show that $\gamma_{H(\rho,E_0,E)}(E)=\langle\theta\rho\rangle_{\rm q}$ in both cases $A$ and $B$.}
$$
\gamma_{H(\rho,E_0,E)}(E)=\langle\theta\rho\rangle_{\rm q}=\left\{\begin{array}{ll}
        \frac{1}{n}\sum_{i=1}^{n}|\varphi_i\rangle\langle \varphi_i|\;&\textit{in case } A \medskip\\
        \frac{e^{-\beta_mH(\rho,E_0,E)}}{\Tr e^{-\beta_m H(\rho,E_0,E)}}\; & \textit{in case } B
        \end{array}\right.
$$
and
\begin{equation}\label{psi-exp}
\begin{array}{c}
 \displaystyle S(\gamma_{H(\rho,E_0,E)}(E))=S\!\left(\langle\theta\rho\rangle_{\rm q}\right)=H\!\left(\langle(\theta p_1,\theta p_2,...)\rangle_{\rm c}\right)\qquad\qquad\qquad\qquad\qquad\\\\ = \left\{\begin{array}{ll}
        \ln n\;&\textit{in case } A \medskip\\
        \eta(1-\theta d_{m})+(1-\theta d_m)\ln m+\theta s_m+d_m\eta(\theta)\; & \textit{in case } B
        \end{array}\right.\!,
\end{array}
\end{equation}
where $H$ is the Shannon entropy, $\,\langle\cdot\rangle_{\rm c}$ and $\,\langle\cdot\rangle_{\rm q}$ are the maps defined in (\ref{psi-c}) and (\ref{psi-q}).}
\end{theorem}\medskip

\begin{remark}\label{main-r+} The sense of case $A$ is simple: it is a "saturation mode", since in this case  the "energy" $\,\Tr H\bar{\rho}\,$  of the state $\,\bar{\rho}\doteq\frac{1}{n}\sum_{i=1}^{n}|\varphi_i\rangle\langle \varphi_i|\,$
does not exceed $E$ for any operator $H$ in $\,\M_E(\rho,E_0)=\M(\rho,E_0)$ (see the below  proof of Theorem \ref{main}).

In case $A$, the operator $H(\rho,E_0,E)$ is not uniquely defined. The choice (\ref{r-c})  of $H(\rho,E_0,E)$ makes the function $E\mapsto H(\rho,E_0,E)$ continuous on $(0,+\infty)$ w.r.t. the operator norm. This can be shown by using the claim of Lemma \ref{ml} before (\ref{Z-def+c}) and by noting that
the $n$-tuple $\{h^B_{i}\}_{i=1}^n$ in (\ref{HB-exp-3}) with $\,m=\tilde{m}\,$ coincides with the $n$-tuple $\{h^A_{i}\}_{i=1}^n$ in (\ref{r-c}).
\end{remark}\smallskip

\begin{remark}\label{main-r+n} The crucial parameter defining the operator $H(\rho,E_0,E)$ in case $B$ is the dimension $m$ of $\ker H(\rho,E_0,E)$ equal to $\mu_{\rm q}(\theta\rho)$. In Section 5 (Examples \ref{one}-\ref{three}) below it is shown how
this parameter changes with increasing $E$ for fixed $\rho$ and $E_0$ (see Fiqures 1-5, green lines).

Note that the condition determining $\,m=\mu_{\rm q}(\theta\rho)\,$ for given $E$ and $E_0$ depends only on the $m+1$ maximal eigenvalues of a state $\rho$.

For given $S(\rho)$ the parameter $s_k$ (introduced in Theorem \ref{main}) is completely determined
by the $k$ maximal eigenvalues of a state $\rho$. This and the previous remark show that
$$
S(\gamma_{H(\rho,E_0,E)}(E))=S(\gamma_{H(\sigma,E_0,E)}(E))
$$
for any state $\sigma$ with the spectral representation $\,\sigma=\sum\limits_{i=1}^{n} q_{i}|\psi_i\rangle\langle \psi_i|\,$ such that
$S(\rho)=S(\sigma)$ and $p_i=q_i$, $i=1,2,...,m+1$, where $m=\dim\ker H(\rho,E_0,E)=\mu_{\rm q}(\theta\rho)$.\smallskip

The above observations  show that \emph{for any infinite rank state $\rho$ with finite entropy $S(\rho)$ the parameters $\beta_m$, $c_m$ and the projector $P_m$ in representation (\ref{HB-exp-2}) of  $H(\rho,E_0,E)$ and the value of $S(\gamma_{H(\rho,E_0,E)}(E))$ can be determined using a finite number of maximal eigenvalues of $\rho$}. This is important for practical applications of Theorem \ref{main}.
\end{remark}\smallskip

\begin{remark}\label{main-r}
If $E=E_0$ then case B holds, $\theta=m=1$ and the unique operator $H(\rho,E_0,E)$ has the following simple form
\begin{equation}\label{HH}
H(\rho,E_0,E_0)=\sum_{i=1}^{n} \frac{E_0(\ln p_1-\ln p_i)}{S(\rho)+\ln p_1}\shs|\varphi_i\rangle\langle \varphi_i|
\end{equation}
(it is assumed that $\D(H(\rho,E_0,E_0))$ is contained in the linear hull of the system $\{\varphi_i\}_{i=1}^{n}$). It follows from (\ref{psi-exp}) that  in this case
\begin{equation}\label{E0}
\gamma_{H(\rho,E_0,E_0)}(E_0)=\rho\quad \textrm{and hence}\quad S(\gamma_{H(\rho,E_0,E_0)}(E_0))=S(\rho).
\end{equation}
\end{remark}

\medskip

\emph{Proof Theorem \ref{main}.} Following the notation used in Section 2 denote $\sup_{\Tr H\sigma\leq E}S(\sigma)$ by $F_H(E)$.\smallskip

The claims of the theorem  concerning case $A$ are proved by simple arguments.
Indeed, assume that $n=\rank\rho<+\infty$, $np_nE\geq E_0$ and $H$ is an arbitrary operator in $\M(\rho,E_0)$ with  representation (\ref{H-form}).
Then
$$
p_n\sum_{k=1}^{+\infty}\sum_{i=1}^{n}h_k|\langle\tau_k|\varphi_i\rangle|^2\leq\sum_{k=1}^{+\infty}\sum_{i=1}^{n}h_kp_i|\langle\tau_k|\varphi_i\rangle|^2=\Tr H\rho \leq E_0
$$
because $p_n$ is the minimal eigenvalue of $\rho$. Hence,
$$
\Tr H\bar{\rho}=\frac{1}{n}\sum_{k=1}^{+\infty}\sum_{i=1}^{n}h_k|\langle\tau_k|\varphi_i\rangle|^2 \leq  \frac{E_0}{np_n} \leq E,
$$
where $\,\bar{\rho}\doteq\frac{1}{n}\sum_{i=1}^{n}|\varphi_i\rangle\langle \varphi_i|$ is the chaotic state on the subspace $\,\supp\rho$. If follows that
the r.h.s. of (\ref{main+}) is not less then $\,S(\bar{\rho})=\ln n$. At the same time, it is not greater than $\ln n$, because $\,F_{H}(E)\leq\ln n\,$ for any  operator $H$ in $\M(\rho,E_0)$ such that $\,\D(H)=\supp\rho\,$ (due to the upper bound $\,\ln n\,$ on the entropy of a state on an $n$-dimensional subspace).

The last claim  of the theorem in case $A$ is proved by noting that in this case $\,\langle\theta\rho\rangle_{\rm q}=\bar{\rho}\,$ is the chaotic state on the subspace $\,\supp\rho$. \smallskip

The claims concerning case $A$ can be also deduced from Lemma \ref{bl} in  Appendix A-1 using the below arguments.
\smallskip

Consider case $B$. Let $H(\rho,E_0,E)$ be the operator defined by one of the equivalent expressions (\ref{HB-exp-1})-(\ref{HB-exp-3}). To prove (\ref{main+}) it suffices to show that
\begin{equation}\label{main++}
\sup_{\Tr H(\rho,E_0,E)\sigma\leq E}S(\sigma)=\inf_{H\in\M(\rho,E_0)}F_H(E)
\end{equation}
and that the operator $H(\rho,E_0,E)$ belongs to the set $\M_E(\rho,E_0)$.\smallskip

Using Courant-Fischer theorem (cf.~\cite[Proposition II-3]{BDJ}) it is easy to show that
for any operator $H$  in $\M(\rho,E_0)$ with representation (\ref{H-form}) the operator
\begin{equation}\label{H-form+}
 H_*=\sum_{k=1}^{+\infty} h_{k}|\varphi_k\rangle\langle \varphi_k|
\end{equation}
(where $\{\varphi_k\}_{k=1}^{+\infty}$ is an extension of the system $\{\varphi_k\}_{k=1}^{n}$ to a basis in $\H$ in the case  $\,n<+\infty$)
belongs to the set  $\M(\rho,E_0)$. Since it is clear that $F_H(E)=F_{H_*}\!(E)$ for any $E>0$, we have
\begin{equation}\label{e-1}
\inf_{H\in\M(\rho,E_0)}F_H(E)=\inf_{H\in\M_*(\rho,E_0)}F_H(E),\quad E>0,
\end{equation}
where $\M_*(\rho,E_0)$ is the subset of $\M(\rho,E_0)$ consisting of operators having form (\ref{H-form+}) with
a nondecreasing sequence $\{h_k\}_{k=1}^{+\infty}\subset[0,+\infty]$ such that $h_1=0$ and $\sum_{k=1}^{+\infty}h_kp_k\leq E_0$.

By Lemma \ref{V} we have\footnote{$g(H)$ is defined before (\ref{Gibbs}).}
$$
F_H(E)=\inf_{\beta\geq0}\left( E\beta +\ln\Tr e^{-\beta H}\right)=\inf_{\beta\geq 0}\left( E\beta+\ln\sum_{k=1}^{+\infty}e^{-\beta h_k}\right)
$$
for any operator $H\in\M(\rho,E_0)$ with the spectrum $\{h_k\}_{k=1}^{+\infty}$ ($h_1=0$).
Hence,
\begin{equation}\label{e-2}
\begin{array}{rl}
\displaystyle\inf_{H\in\M_*(\rho,E_0)}F_H(E)\;\,=&\displaystyle\inf_{H\in\M_*(\rho,E_0)}\inf_{\beta\geq0}\left( E\beta+\ln\Tr e^{-\beta H}\right)\\\\
=& \displaystyle\inf_{\beta\geq0}\inf_{H\in\M_*(\rho,E_0)}\left( E\beta+\ln\Tr e^{-\beta H}\right)\\\\
=& \displaystyle\inf_{\beta\geq0}\left( E\beta+ \inf_{\{h_k\}_{k=1}^n\in\mathfrak{H}(\rho)}\ln\sum_{k=1}^ne^{-\beta h_k}\right),
\end{array}
\end{equation}
where $\mathfrak{H}(\rho)$ is the set of $n$-tuples $\{h_k\}_{k=1}^n$  of nonnegative numbers arranged in the nondecreasing order such that $\sum_{k=1}^nh_kp_k=E_0$ and $h_1=0$.

Since $F_{cH}(cE)=F_{H}(E)$ for any $c>0$, it suffices to consider the case $E_0=1$. So,
expression (\ref{e-1}) and (\ref{e-2}) allow us to prove (\ref{main+}) by applying Lemma \ref{bl} in  Appendix A-1
with $\{a_i\}_{i=1}^n=\{p_i\}_{i=1}^n$, $b=\beta$, $\theta=E\,$ and $\,c=0\,$ by noting that in this case $d_0=\Tr\rho=1$ and $s_0=H(\{p_i\}_{i=1}^n)=S(\rho)<+\infty$.     \smallskip

To show that the operator $H(\rho,E_0,E)$ belongs to the set $\M_E(\rho,E_0)$ in the case $\rank \rho=+\infty$ it suffices, by Lemma \ref{Gibbs-r} in Section 2, to note that
Lemma \ref{bl} in  Appendix A-1 guarantees the validity of equality (\ref{imp-in}) with $H=H(\rho,E_0,E)$ and $\beta=\beta_m$.

The expression $\,\gamma_{H(\rho,E_0,E)}(E)=\langle\theta\rho\rangle_{\rm q}\,$ in case $B$ also follows from Lemma \ref{bl} by the above arguments. $\Box$\medskip

\section{Properties of the function $E\mapsto S(\gamma_{H(\rho,E_0,E)}(E))$}

Keeping in mind the applications of Theorem \ref{main} described in Section 6 we have to analyse
properties of the function $E\mapsto S(\gamma_{H(\rho,E_0,E)}(E))$ on $\mathbb{R}_+$, especially, its behavior for $E\gg E_0$.
Theorem \ref{main} implies that
\begin{equation}\label{psi-exp+}
S(\gamma_{H(\rho,E_0,E)}(E))=S\!\left(\langle\theta\rho\rangle_{\rm q}\right)=H\!\left(\langle(\theta p_1,\theta p_2,...)\rangle_{\rm c}\right),
\end{equation}
where $\{p_1, p_2,...\}$ is the spectrum of $\rho$ (arranged in the non-increasing order), $H$ is the Shannon entropy, $\langle\cdot\rangle_{\rm c}$ and $\langle\cdot\rangle_{\rm q}$ are the maps defined in (\ref{psi-c}) and (\ref{psi-q}).\smallskip\pagebreak

Denote the function $E\mapsto S(\gamma_{H(\rho,E_0,E)}(E))$ by $G_{\!E_0}^{\shs\rho}(E)$. \smallskip

\begin{theorem}\label{main-p} \emph{Let the assumptions and notation of Theorem \ref{main} be valid.}\smallskip

\emph{For an arbitrary mixed state $\rho$ the function $G_{\!E_0}^{\shs\rho}(E)$ is nondecreasing concave and continuously differentiable\footnote{One can show that the second derivative of $G_{\!E_0}^{\shs\rho}(E)$ is not continuous at the points $E_2,E_3,...$ defined below.} on $(0,+\infty)$.  Moreover, $G_{\!E_0}^{\shs\rho}(E_0)=S(\rho)$,}
\begin{equation}\label{G-lr}
\!\!\!\lim_{E\to 0^+} G_{\!E_0}^{\shs\rho}(E)=0,\quad  \lim_{E\to +\infty} G_{\!E_0}^{\shs\rho}(E)=\left\{\begin{array}{ll}
        \ln n&\textrm{if}\;\; n<+\infty \\
        +\infty & \textrm{if}\;\; n=+\infty \\
        \end{array}\right.\!, \quad  \lim_{E\to  +\infty} [G_{\!E_0}^{\shs\rho}]'(E)=0.
\end{equation}

\smallskip

\emph{If $\,n<+\infty\,$ then the function $G_{\!E_0}^{\shs\rho}(E)$ can be represented as\footnote{$\mathbf{1}_{B}$ is the indicator function of a set $B$. The r.h.s. of (\ref{G-1}) and (\ref{G-2}) are well defined by Remark \ref{main-r++}.}
\begin{equation}\label{G-1}
G_{\!E_0}^{\shs\rho}(E)=\sum_{m=1}^{n-1} \mathbf{1}_{[E_m,E_{m+1})}(E)S_m(E)+\mathbf{1}_{[E_n,+\infty)}(E)\ln n.
\end{equation}
where
$$
S_m(E)=H(\underbrace{c,...,c}_{m\textrm{ items}},\theta p_{m+1},\theta p_{m+2},...)=\theta s_m+\eta(1-d_m\theta)+(1-d_m\theta)\ln m + d_m\eta(\theta),
$$
$\theta=E/E_0$, $E_1\doteq0$ and $\,E_m\doteq\frac{E_0}{d_m+mp_m}$, $m\in\N\cap(1,n],$ $d_m$ and $s_m$ are the parameters defined in Theorem \ref{main}, $c=(1-\theta d_m)/m$, $H$ is the Shannon entropy, $\eta(x)=-x\ln x$.}\smallskip

\emph{If $\,n=+\infty\,$ then the function $G_{\!E_0}^{\shs\rho}(E)$ can be represented as
\begin{equation}\label{G-2}
G_{\!E_0}^{\shs\rho}(E)=\sum_{m=1}^{+\infty} \mathbf{1}_{[E_m,E_{m+1})}(E)S_m(E),
\end{equation}
where $S_m(E)$ is defined before.}

\emph{For any $\,n\leq+\infty\,$ and $\,m\in\N\cap(1,n]\,$ the following relations hold
\begin{equation}\label{FR}
G_{\!E_0}^{\shs\rho}(E_m)=E_m(s_m+m\eta(p_m))-\ln E_m,\qquad [G_{\!E_0}^{\shs\rho}]'(E_m)=s_m+d_m\ln p_m.
\end{equation}}
\end{theorem}\medskip

\begin{remark}\label{main-r++}
It is easy to see that $E_m\leq E_{m+1}$ for any $m$ in $\N\cap[1,n)$. If $n<+\infty$ then the union of disjoint sets\footnote{Some of these sets may be empty, since it is easy to see that $E_m=E_{m+1}$ if $p_m=p_{m+1}$, $m>1$.} $[E_m,E_{m+1})$, $m\in\N\cap[1,n)$, and  $[E_n,+\infty)$ coincides with $[0,+\infty)$. If $n=+\infty$ then $E_m\to+\infty$ as $m\to+\infty$ and, hence, the union of disjoint sets $[E_m,E_{m+1})$, $m\in\N$, also coincides with $[0,+\infty)$.
\end{remark}\smallskip

\emph{Proof.} Since $G_{\!E_0}^{\shs\rho}(E)$ depends only on $\theta\doteq E/E_0$, to simplify notation we may assume in what follows that
$E_0=1$. For natural $k$ denote $\N\cap[1,k+1)$ by $I_k$. \smallskip

It is easy to show that the condition $E\in [E_m,E_{m+1})$  (resp. $E\in [E_n,+\infty)$ in the case $n<+\infty$) holds if and only if $\,m=\mu_{\rm q}(E\rho)$
(resp. $n=\mu_{\rm q}(E\rho)$). So, the expressions (\ref{G-1}) and (\ref{G-2}) directly follow from Theorem \ref{main}.

To prove continuity of the functions $G_{\!E_0}^{\shs\rho}$ and $[G_{\!E_0}^{\shs\rho}]'$ on $(0,+\infty)$ it suffices to
show continuity  of these functions at the gluing points $E_2,E_3,...$ taking into account that some of these points may coincide (as $p_m=p_{m+1}$ for $m>1$ implies that $E_m=E_{m+1}$ and vise versa).

Assume that $p_{m-j+1}=...=p_{m}>p_{m+1}$ for some $m\in I_{n-1}$ and $j\leq m$. In the case $j<m$, assume also that $p_{m-j}>p_{m-j+1}$.
Then
$E_{t}<E_{t+1}=...=E_{m}<E_{m+1}$, where $t=1\,$ in the case $j=m$ and $\,t=m-j$ otherwise. By noting that
\begin{equation}\label{sk-exp}
S_m(E)=E\!\left(s_m+\eta(E^{-1}-d_m)+(E^{-1}-d_m)\ln m -\eta(E^{-1})\right),
\end{equation}
$E^{-1}_m-d_{m-j}=(m-j)p_{m-j+1}=(m-j)p_{m}\,$ and  $\,E^{-1}_m-d_m=mp_m$ it is easy to show that$$
S_{t}(E_m)=S_{m}(E_m)=E_m(s_m+m\eta(p_m))-\ln E_m,\qquad S'_{t}(E_m)=S'_{m}(E_m)=s_m+d_m\ln p_m.
$$
These equalities imply the expressions in  (\ref{FR}).\smallskip

To prove continuity of the functions $G_{\!E_0}^{\shs\rho}$ and $[G_{\!E_0}^{\shs\rho}]'$ at the point $E_n$  in the case $n<+\infty$ assume that
$p_{m+1}=...=p_{n}$ for some $m\in\{0\}\cap I_{n-1}$ and that $p_{m}>p_{m+1}$ in the case $m>0$. Then $E_{t}<E_{t+1}=...=E_{n}=\frac{1}{n p_n}$, where $t=1$ the case $m=0$ and $t=m$ otherwise. We have to show that
$\,S_{t}(E_n)=\ln n\,$ and $\,S'_{t}(E_n)=0$. These equalities can be easily verified  by using (\ref{sk-exp}) and by noting that
$E^{-1}_n-d_{t}=n p_n-p_n(n-t)=tp_n$.

By the continuity of $[G_{\!E_0}^{\shs\rho}]'$ to prove the concavity of $G_{\!E_0}^{\shs\rho}$ it suffices to
show that $[G_{\!E_0}^{\shs\rho}]''(E)\leq 0$ in any nonempty interval $(E_{m},E_{m+1})$, $m\in I_{n-1}$. Direct calculation shows that
$$
[G_{\!E_0}^{\shs\rho}]''(E)=S''_m(E)=-\frac{d_m}{E-E^2d_m}<0\qquad  \forall E\in (E_{m},E_{m+1}).
$$

The limit  relations in  (\ref{G-lr}) follows from the above properties of the functions $G_{\!E_0}^{\shs\rho}$ and $[G_{\!E_0}^{\shs\rho}]'$
and the expression in  (\ref{FR}).

The equality $G_{\!E_0}^{\shs\rho}(E_0)=S(\rho)$ is mentioned in Remark \ref{main-r}. $\Box$ \medskip

\begin{corollary}\label{main-c} \emph{Let $\rho$ be an arbitrary mixed state with finite entropy. Then}\smallskip
\begin{equation}\label{H-cond-2}
 \lim_{E\to +\infty} (1/E)G_{\!E_0}^{\shs\rho}(E)=\lim_{E\to+\infty}(1/E)F_{H(\rho,E_0,E)}(E)=0.
\end{equation}
\end{corollary}

\emph{Proof.} The claim of the lemma follows  directly from the last limit relation in (\ref{G-lr}) by L'Hopital's rule.

Another way to prove it is to note that for any state $\rho$ with finite entropy
there is a positive operator $H$ satisfying conditions (\ref{H-cond}) and (\ref{star}) such that $\Tr H\rho=E_0>0$ (see, f.i., Proposition 4 in \cite{EC}) and to use the equivalence of (\ref{H-cond}) and (\ref{H-cond-a}), since the definition of $H(\rho,E_0,E)$ implies that
$$
G_{\!E_0}^{\shs\rho}(E)\doteq S(\gamma_{H(\rho,E_0,E)}(E))\leq S(\gamma_{H}(E))
$$
for any positive operator $H$ satisfying conditions (\ref{H-cond}) and (\ref{star}) such that $\Tr H\rho\leq 1$. $\Box$\smallskip

\begin{remark}\label{main-r++n}
Hamiltonians $H$ of real quantum systems satisfy the  Gibbs condition (\ref{H-cond}) (cf.\cite{W})
which is stronger than condition (\ref{H-cond-w}). As mentioned in Section 2, condition (\ref{H-cond})
is equivalent to the asymptotic property (\ref{H-cond-a}) of the function $F_{H}(E)$.

It is easy to see that the "optimal" operator $H(\rho,E_0,E)$ satisfies the Gibbs condition (\ref{H-cond})
if and only if $\rho$ is such a state that $\,\Tr \rho^{\shs\beta}<+\infty\,$ of all $\beta>0$. So, if $S(\rho)<+\infty$ but the last condition
does not hold\footnote{Example of such a state $\rho$ can be found in \cite{EC} (before Proposition 2).} then
\begin{equation}\label{H-cond-1}
 \lim_{E\to +\infty} (1/E)S(\gamma_{H(\rho,E_0,E')}(E))=\lim_{E\to +\infty}(1/E)F_{H(\rho,E_0,E')}(E)>0
\end{equation}
for any given $E_0$ and $E'$. At the same time, Corollary \ref{main-c} implies that relation (\ref{H-cond-2})
holds for \emph{arbitrary} state $\rho$ with finite entropy. The difference between (\ref{H-cond-1}) and (\ref{H-cond-2}) is clear: in the first case
$H(\rho,E_0,E')$ is a fixed operator, in the second one the operator $H(\rho,E_0,E)$ adapts to the current value of $E$.
\end{remark}

\section{Examples}

In this section  we describe the "optimal" operator $H(\rho,E_0,E)$ and the state $\gamma_{H(\rho,E_0,E)}(E)\\=\left\langle\frac{E}{E_0}\rho\right\rangle_{\rm q}$ pointing attention to analysis  of the functions  $\,E\mapsto S(\gamma_{H(\rho,E_0,E)}(E))\,$ and  $\,E\mapsto \dim\ker H(\rho,E_0,E)=\mu_{\rm q}\!\left(\frac{E}{E_0}\rho\right)$ for several quantum states $\rho$ with finite entropy.\smallskip

\begin{example}\label{qubit}
Assume that $\,\rho_2=p_1\varphi_1+p_2\varphi_2\,$  ($\varphi_i= |\varphi_i\rangle\langle \varphi_i|$, $i=1,2$) is a state of rank $2$, where $p_1\geq p_2>0$. In this case
$d_1=p_2\,$ and $\,E_2=\frac{E_0}{2p_2}$. Theorem \ref{main-p} shows that
$$
\gamma_{H(\rho_2,E_0,E)}(E)=\langle\theta \rho_2\rangle_{\rm q}=\left\{\begin{array}{ll}
        \!(1-\theta p_2)\varphi_1+\theta p_2\varphi_2   &\textrm{if}\;\;  E<\frac{E_0}{2p_2}\\\\
        \!\frac{1}{2}(\varphi_1+\varphi_2) & \textrm{if}\;\;  E\geq \frac{E_0}{2p_2}
        \end{array}\right.\!\!,
$$
$$
S(\gamma_{H(\rho_2,E_0,E)}(E))=\left\{\begin{array}{ll}
        h(\theta p_2)  &\textrm{if}\;\;  E<\frac{E_0}{2p_2} \\\\
        \ln2& \textrm{if}\;\;  E\geq \frac{E_0}{2p_2}
        \end{array}\right.\!\!,
$$
where $\,\theta=E/E_0$  and $h$ is the binary entropy. The first region of $E$ corresponds to case $B$, the second one -- to case $A$.  The "optimal" operator has the form
$$
H(\rho_2,E_0,E)=\left\{\begin{array}{ll}
        \!0\shs\varphi_1+\frac{E_0}{p_2}\shs\varphi_2& \textrm{if}\;\;  E<\frac{E_0}{2p_2}\\\\
        \!h_1\shs\varphi_1+h_2\varphi_2& \textrm{if}\;\; E\geq \frac{E_0}{2p_2}
        \end{array}\right.\!\!,
$$
where $\,h_1\,$ and $\,h_2\,$ is any nonnegative numbers such that $h_1h_2=0$, $\,h_1p_1+h_2p_2\leq E_0\,$ and it is assumed that $\D(H(\rho_2,E_0,E))$ is the linear span of the vectors  $|\varphi_1\rangle$ and  $|\varphi_2\rangle$.
\end{example}\smallskip

\begin{example}\label{quthreet} Assume that $\,\rho_3=p_1\varphi_1+p_2\varphi_2+p_3\varphi_3\,$ ($\varphi_i= |\varphi_i\rangle\langle \varphi_i|$, $i=1,2,3$) is a state of rank $3$, where $p_1\geq p_2\geq p_3>0$. In this case
$\,d_1=p_2+p_3$, $d_2=p_3$, $\,E_2=\frac{E_0}{2p_2+p_3}$ and $\,E_3=\frac{E_0}{3p_3}$ . Theorem \ref{main-p} shows that
$$
\gamma_{H(\rho_3,E_0,E)}(E)=\langle\theta \rho_3\rangle_{\rm q}=\left\{\begin{array}{ll}
        \!(1-\theta p_2-\theta p_3)\varphi_1+\theta p_2\varphi_2+\theta p_3\varphi_3   &\textrm{if}\;\;  E<\frac{E_0}{2p_2+p_3}\\\\
        \!\frac{1-\theta p_3}{2}(\varphi_1+\varphi_2)+\theta p_3\varphi_3& \textrm{if}\;\;  \frac{E_0}{2p_2+p_3}\leq E< \frac{E_0}{3p_3}\\\\
        \!\frac{1}{3}(\varphi_1+\varphi_2+\varphi_3)& \textrm{if}\;\;  E\geq\frac{E_0}{3p_3}
        \end{array}\right.\!\!,
$$
$$
S(\gamma_{H(\rho_3,E_0,E)}(E))=\left\{\begin{array}{ll}
        \eta(1-\theta p_2-\theta p_3)+\eta(\theta p_2\varphi_2)+\eta(\theta p_3\varphi_3)  &\textrm{if}\;\;  E<\frac{E_0}{2p_2+p_3}\\\\
        \!2\eta\!\left(\frac{1-\theta p_3}{2}\right)+\eta(\theta p_3)& \textrm{if}\;\;  \frac{E_0}{2p_2+p_3}\leq E< \frac{E_0}{3p_3}\\\\
        \!\ln 3& \textrm{if}\;\;  E\geq \frac{E_0}{3p_3}
        \end{array}\right.\!\!,
$$
where $\,\theta=E/E_0$  and $\eta(x)=-x\ln x$. The first two regions of $E$ correspond to case $B$, the third one -- to case $A$. The "optimal" operator has the form
$$
H(\rho_3,E_0,E)=\left\{\begin{array}{ll}
        \!C\!\left(0\shs\varphi_1+\left(\ln\frac{1-\theta p_2-\theta p_3}{\theta p_2}\right)\varphi_2+\left(\ln\frac{1-\theta p_2-\theta p_3}{\theta p_3}\right)\varphi_3\right)  &\textrm{if}\;\;  E<\frac{E_0}{2p_2+p_3}\\\\
        \!0\shs\varphi_1+0\shs\varphi_2+\frac{E_0}{p_3}\shs\varphi_3& \textrm{if}\;\;  \frac{E_0}{2p_2+p_3}\leq E< \frac{E_0}{3p_3}\\\\
        \!h_1\shs\varphi_1+h_2\varphi_2+h_3\varphi_3& \textrm{if}\;\;  E\geq \frac{E_0}{3p_3}
        \end{array}\right.\!\!,
$$
where   $C=E_0\left(p_2\ln\frac{1-\theta p_2-\theta p_3}{\theta p_2}+p_3\ln\frac{1-\theta p_2-\theta p_3}{\theta p_3}\right)^{-1}$, $\,h_1,$ $\,h_2\,$ and $\,h_3\,$ are any nonnegative numbers such that $\,h_1h_2h_3=0$, $\,h_1p_1+h_2p_2+h_3p_3\leq E_0\,$ and it is assumed that $\D(H(\rho_3,E_0,E))$ is the linear span of the vectors  $|\varphi_1\rangle$, $|\varphi_2\rangle$ and  $|\varphi_3\rangle$.
\end{example}\smallskip

\begin{example}\label{one} Assume that $\rho=\bar{\rho}_n\doteq\frac{1}{n}\sum_{i=1}^{n}\varphi_i$ ($\varphi_i= |\varphi_i\rangle\langle \varphi_i|$, $i=\overline{1,n}$) for some $n<+\infty$. In this case $p_i=\frac{1}{n}$ for all $i=1,2,..,n$. Hence, $d_k=\frac{n-k}{n}$, $k=1,2,..,n,$ and  $E_2=E_3=...=E_n=E_0$. So, Theorem \ref{main-p} shows that
$$
\gamma_{H(\bar{\rho}_n,E_0,E)}(E)=\langle\theta \bar{\rho}_n\rangle_{\rm q}=\left\{\begin{array}{ll}
        \!\left(1-\frac{\theta(n-1)}{n}\right)\!\varphi_1+\frac{\theta}{n}\shs\varphi_2+...+\frac{\theta}{n}\shs\varphi_n   &\textrm{if}\;\;  E< E_0\\\\
        \!\frac{1}{n}(\varphi_1+\varphi_2+...+\varphi_n)& \textrm{if}\;\;  E\geq E_0
        \end{array}\right.\!\!,
$$
$$
S(\gamma_{H(\bar{\rho}_n,E_0,E)}(E))=\left\{\begin{array}{ll}
        \!\eta\!\left(1-\frac{\theta(n-1)}{n}\right)+(n-1)\eta\!\left(\frac{\theta}{n}\right)   &\textrm{if}\;\;  E< E_0\\\\
        \!\ln n& \textrm{if}\;\;  E\geq E_0
        \end{array}\right.\!\!,
$$
where $\,\theta=E/E_0\,$ and $\,\eta(x)=-x\ln x$. The  first region of  $E$ corresponds to case $B$, the second one -- to case $A$. In both cases, the "optimal" operator has the form
$$
H(\bar{\rho}_n,E_0,E)=0\shs\varphi_1+\frac{E_0n}{(n-1)}
        (\varphi_2+\varphi_3+...+\varphi_n),
$$
where it is assumed that $\D(H(\bar{\rho}_n,E_0,E))$ is the linear span of the vectors  $|\varphi_1\rangle$,...,$|\varphi_n\rangle$.\footnote{In case $A$ the operator $H(\bar{\rho}_n,E_0,E)$ is not unique. The presented version of $H(\bar{\rho}_n,E_0,E)$ for $E\geq E_n$ corresponds to the choice (\ref{r-c}) of this operator which makes the function $E\mapsto H(\bar{\rho}_n,E_0,E)$ continuous on $(0,+\infty)$ w.r.t. the operator norm.}

In Fig.1, the graph of the function $\,(E/E_0)\mapsto S(\gamma_{H(\bar{\rho}_n,E_0,E)}(E))\,$ with $\,n=10\,$ is shown
along with the values of the parameter $m=\mu_{\rm q}\!\left(\frac{E}{E_0}\bar{\rho}_n\right)$ -- the dimension of the null subspace of $H(\bar{\rho}_n,E_0,E)$. The coincidence of $S(\gamma_{H(\bar{\rho}_n,E_0,E)}(E))$ with $S(\bar{\rho}_n)$ at the point $E/E_0=1$ agrees with (\ref{E0}).
\end{example}\smallskip

\begin{figure}[t]

\centering

\begin{center}

\includegraphics[scale=0.5, bb=400  450 500 550]{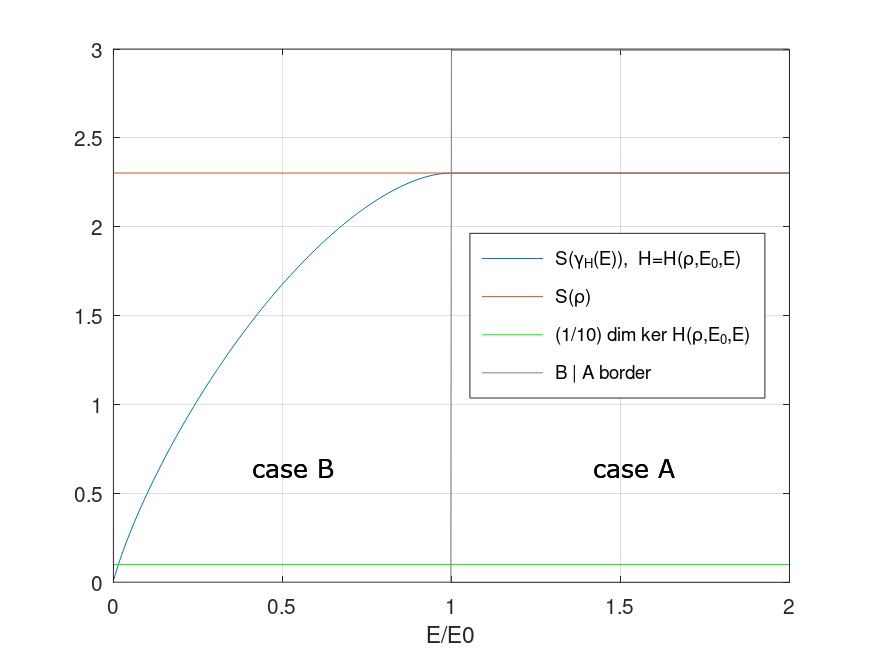}

\vspace{225pt}
\caption{The values of $S(\gamma_{H(\bar{\rho}_n,E_0,E)}(E))$ with $n=10$ along with the values of $m(E)\doteq\dim\ker H(\rho_n,E_0,E)$ with factor $1/10$.}
\end{center}

\label{Fig1n}

\end{figure}

\begin{figure}[t]

\centering

\begin{center}

\includegraphics[scale=0.5, bb=400  450 500 550]{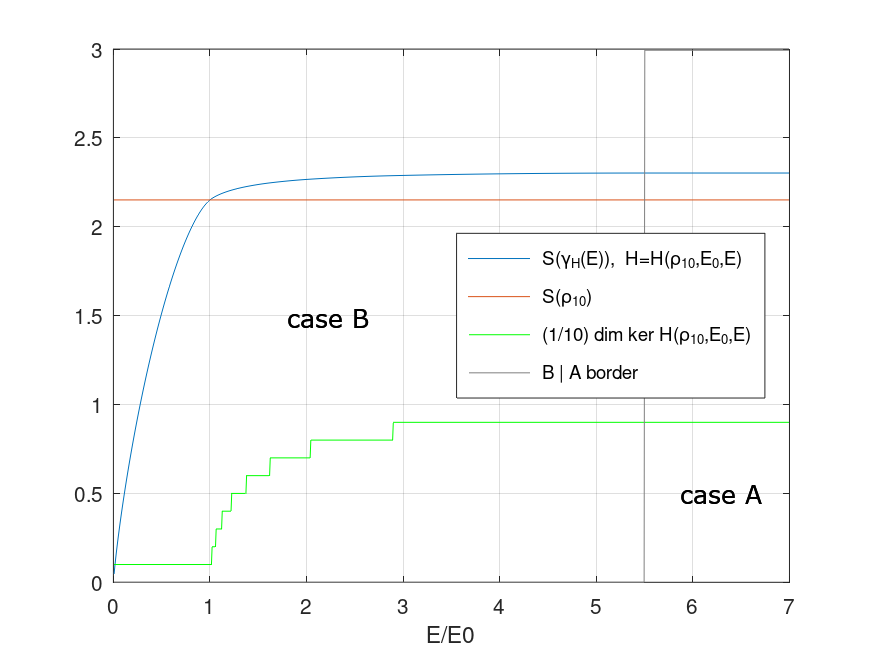}

\vspace{225pt}
\caption{The values of $S(\gamma_{H(\rho_n,E_0,E)}(E))$ with $n=10$ along with the values of $m(E)\doteq\dim\ker H(\rho_n,E_0,E)$ with factor $1/10$.}
\end{center}

\label{Fig2}

\end{figure}

\begin{figure}[t]

\centering

\begin{center}

\includegraphics[scale=0.5, bb=400  450 500 550]{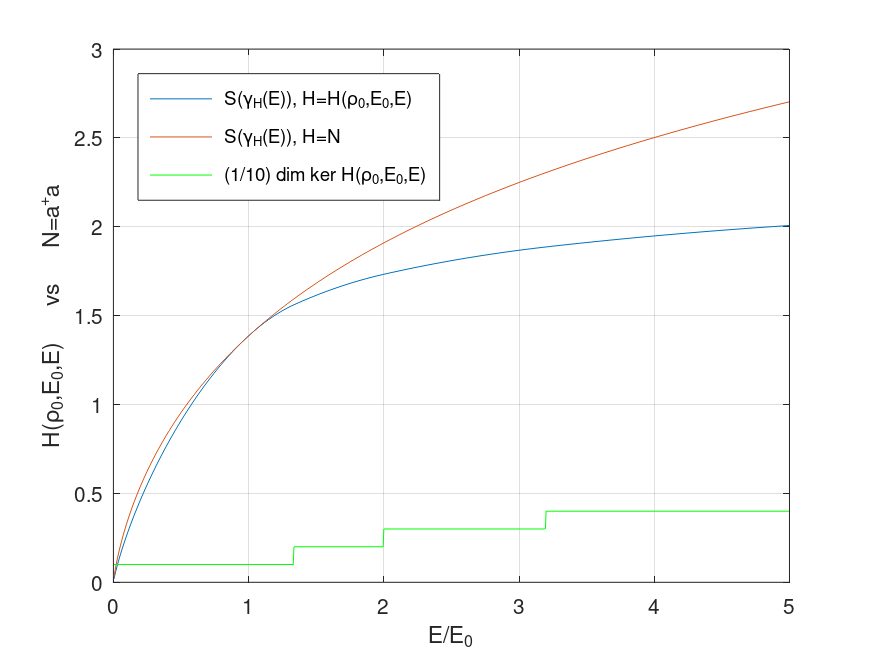}

\vspace{225pt}
\caption{The values of $S(\gamma_{H(\rho_0,E_0,E)}(E))$ and $S(\gamma_N(E))$ with $E_0=1$ along with the values of $m(E)\doteq\dim\ker H(\rho_0,E_0,E)$ with factor $1/10$.}
\end{center}

\label{Fig3}

\end{figure}

\begin{figure}[tt]

\centering

\begin{center}

\includegraphics[scale=0.5, bb=400  450 500 550]{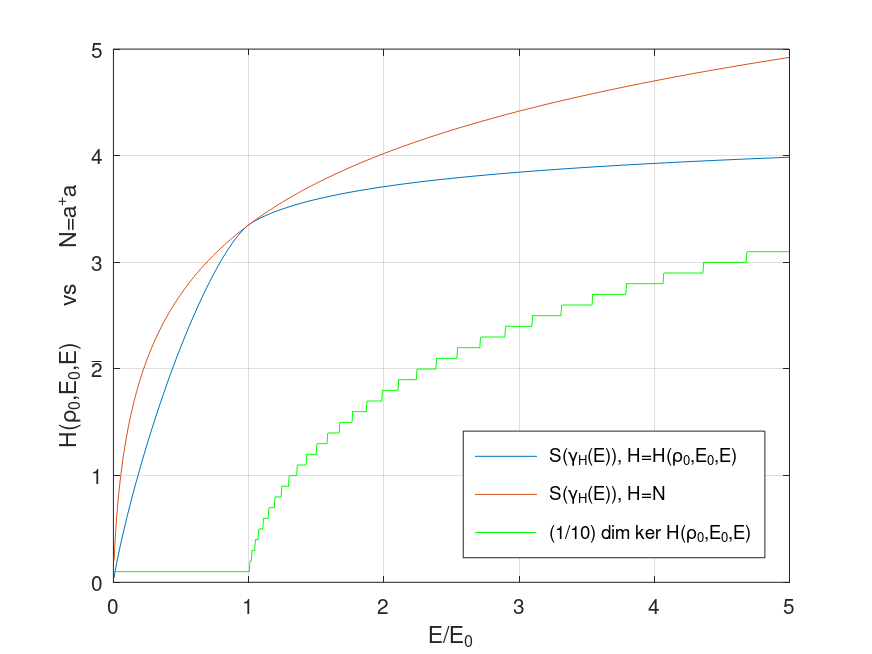}

\vspace{225pt}
\caption{The values of $S(\gamma_{H(\rho_0,E_0,E)}(E))$ and $S(\gamma_N(E))$ with $E_0=10$ along with the values of $m(E)\doteq\dim\ker H(\rho_0,E_0,E)$ with factor $1/10$.}
\end{center}

\label{Fig4}
\end{figure}

\begin{figure}[ttt]

\centering

\begin{center}

\includegraphics[scale=0.5, bb=400  450 500 550]{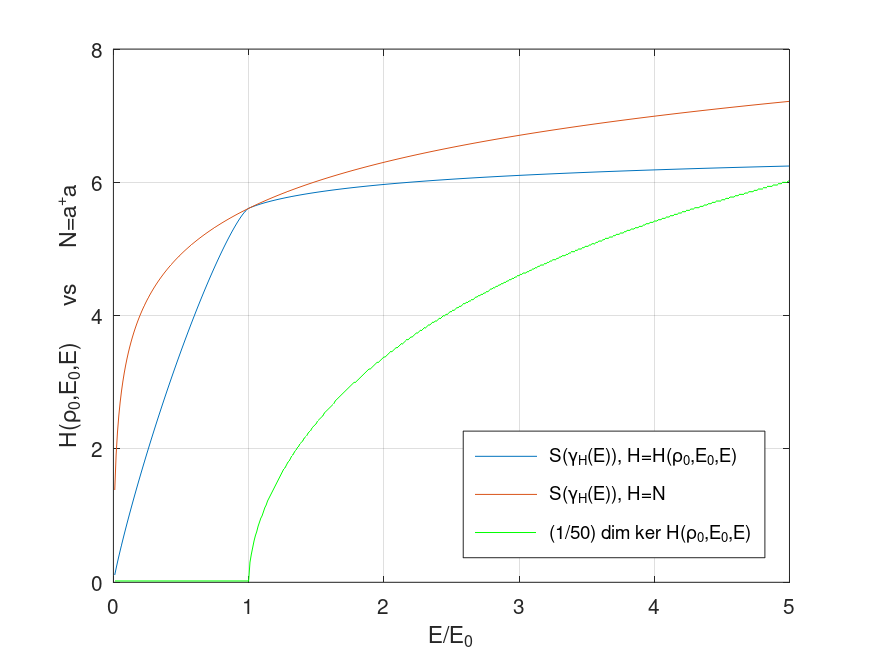}

\vspace{225pt}
\caption{The values of $S(\gamma_{H(\rho_0,E_0,E)}(E))$ and $S(\gamma_N(E))$ with $E_0=100$ along with the values of $m(E)\doteq\dim\ker H(\rho_0,E_0,E)$ with factor $1/50$.}
\end{center}

\label{Fig5}

\end{figure}

\medskip

\begin{example}\label{two} Assume that $\rho_n$ is a finite rank state with lineally decreasing spectrum, i.e. $\,\rho_n=\sum_{i=1}^{n}\tilde{p}_i\varphi_i$ ($\varphi_i= |\varphi_i\rangle\langle \varphi_i|$, $i=\overline{1,n}$), where $\,n<+\infty\,$ and $\,\tilde{p}_i=\frac{2(n-i+1)}{n(n+1)},$  $\,i=1,2,...,n$. In this case $\,\tilde{d}_k=\frac{(n-k)(n-k+1)}{n(n+1)}\,$ and, hence,
$E_k=\frac{E_0n(n+1)}{(n+k)(n-k+1)}$, $k=2,3,..,n$. So, Theorem \ref{main-p} shows that
$$
\!\gamma_{H(\rho_n,E_0,E)}(E)=\langle\theta \rho_n\rangle_{\rm q}=\!\left\{\begin{array}{ll}
        \!c_m\left(\sum_{i=1}^m\varphi_i\right)+\theta \sum_{i=m+1}^n \tilde{p}_{i}\varphi_{i}\! &\textrm{if}\;\,  E\in[E_m,E_{m+1}),\; m<n\\\\
        \!\frac{1}{n}(\varphi_1+\varphi_2+...+\varphi_n)& \textrm{if}\;\,  E\in[E_n,+\infty)
        \end{array}\right.\!\!,
$$

$$
S(\gamma_{H(\rho_n,E_0,E)}(E))=\left\{\begin{array}{ll}
        \! m\eta(c_m)+\sum_{i=m+1}^n\eta(\theta\tilde{p}_{i})   &\textrm{if}\;\,  E\in[E_m,E_{m+1}),\; m<n\\\\
        \!\ln n& \textrm{if}\;\,  E\in[E_n,+\infty)
        \end{array}\right.\!\!,
$$
where $\,\theta=E/E_0$, $\,c_m=\left(1-\theta \sum_{i=m+1}^n \tilde{p}_{i}\right)/m=(1-\theta\tilde{d}_k)/m\,$ and $\,\eta(x)=-x\ln x$. The intervals $[E_m,E_{m+1}),\; m<n$ correspond to case $B$, the interval $[E_n,+\infty)$ -- to case $A$. The "optimal" operator has the form
$$
H(\rho_n,E_0,E)=\left\{\begin{array}{ll}
        \!0\shs\sum_{i=1}^m\varphi_i+\sum_{i=m+1}^n \tilde{h}_i\varphi_i &\textrm{if}\;\,  E\in[E_m,E_{m+1}),\; m<n\\\\
        \!0\shs\sum_{i=1}^{n-1}\varphi_i+\frac{1}{2}E_0n(n+1)\shs\varphi_n& \textrm{if}\;\,  E\in[E_n,+\infty)
        \end{array}\right.\!\!,
$$
where $\,\tilde{h}_i=\frac{E_0(\ln(1-\theta \tilde{d}_m)-\ln(\tilde{p}_im\theta))}{\tilde{s}_m+\tilde{d}_m(\ln(1- \theta \tilde{d}_m)-\ln (m\theta))},$ $i=m+1,..,n$, $\,\tilde{s}_m=\sum_{i=m+1}^n\eta(\tilde{p}_i)\,$ and it is assumed that $\D(H(\rho_n,E_0,E))$ is the linear span of the vectors  $|\varphi_1\rangle$, ...,$|\varphi_n\rangle$.\footnote{In case $A$ the operator $H(\rho_n,E_0,E)$ is not unique. The presented version of $H(\rho_n,E_0,E)$ for $E\geq E_n$ corresponds to the choice (\ref{r-c}) of this operator which makes the function $E\mapsto H(\rho_n,E_0,E)$ continuous on $(0,+\infty)$ w.r.t. the operator norm.}

In Fig.2, the graph of the function $\,(E/E_0)\mapsto S(\gamma_{H(\rho_n,E_0,E)}(E))\,$ with $\,n=10\,$ is shown
along with the values of the parameter $\,m=\mu_{\rm q}\!\left(\frac{E}{E_0}\rho_n\right)$ -- the dimension of the null subspace of $H(\rho_n,E_0,E)$. The coincidence of $S(\gamma_{H(\rho_n,E_0,E)}(E))$ with $S(\rho_n)$ at the point $E/E_0=1$ agrees with (\ref{E0}).
\end{example}
\smallskip

\begin{example}\label{three} Assume that
$$
\rho_0=\gamma_N(E_0)=(1-q)\sum_{i=1}^{+\infty} q^{i-1}|\phi_i\rangle\langle \phi_i|,\qquad q=\frac{E_0}{E_0+1},
$$
is the Gibbs state of a quantum oscillator  corresponding to the mean number of quanta $E_0$,
where $\,N=a^{\dag}a\,$ is the number operator and $\,\{\phi_i\}_{i=1}^{+\infty}\,$ is the Fock basis in $\H$ \cite{H-SCI}. In this case $\,d_k=q^k$,  $\,E_k=E_0/(q^{k-1}(k-q(k-1)),$ $k=2,3,..$, and $\,s_k=q^k(g(E_0)-k\ln q)$, $k\in\N$, where
$\,g(E_0)\doteq(E_0+1)\ln(E_0+1)-E_0\ln(E_0)=S(\rho_0)$.
Hence,  Theorem \ref{main-p} implies that
$$
\!\gamma_{H(\rho_0,E_0,E)}(E)=\langle\theta \rho_0\rangle_{\rm q}=\sum_{m=1}^{+\infty}\mathbf{1}_{[E_m,E_{m+1})}(E)\!\left( \frac{1-\theta q^m}{m}
\sum_{i=1}^m\phi_i+\theta(1-q)\!\!\sum_{i=m+1}^n q^{i-1}\phi_{i}\right).
$$
$$
\!S(\gamma_{H(\rho_0,E_0,E)}(E))=\sum_{m=1}^{+\infty} \mathbf{1}_{[E_m,E_{m+1})}(E)\!\left(\! m\eta\!\left(\frac{1-\theta q^m}{m}\right)+\theta q^m(g(E_0)-m\ln q)+q^m\eta(\theta)\!\right),
$$
where $\,\theta=E/E_0\,$ and $\,\eta(x)=-x\ln x$. The "optimal" operator has the form
$$
H(\rho_n,E_0,E)=\sum_{m=1}^{+\infty}\mathbf{1}_{(E_m,E_{m+1}]}(E)\!\left(0\shs\sum_{i=1}^m\phi_i+\sum_{i=m+1}^{+\infty}(a_m (i-1)+b_{m})\,\phi_i\right),
$$
where $\;a_{m}=\frac{E_0(-\ln q)}{q^m(g(E_0)+\ln(1-q^m\theta)-\ln (q^m\theta m))}\;$ and $\;b_m=\frac{E_0(\ln(1-q^m\theta)-\ln((1-q)\theta m))}{q^m(g(E_0)+\ln(1-q^m\theta)-\ln (q^m\theta m))}$.\medskip

Since $\Tr N\rho_0=E_0$, it is natural to compare the values of $S(\gamma_{H(\rho_0,E_0,E)}(E))$
with the values of $S(\gamma_N(E))=g(E)$. In Fig.3-5, the graphs of the functions $(E/E_0)\mapsto S(\gamma_{H(\rho_0,E_0,E)}(E))$ and $(E/E_0)\mapsto S(\gamma_N(E))$ are shown for different values of $E_0$
along with the values of the parameter $m=\mu_{\rm q}\!\left(\frac{E}{E_0}\rho_0\right)$ -- the dimension of the null subspace of $H(\rho_0,E_0,E)$. The coincidence of $S(\gamma_N(E))$  and $S(\gamma_{H(\rho_0,E_0,E)}(E))$ in the point $E/E_0=1$ in all the figures is due to the equality (\ref{E0}), since in this case $S(\rho_0)=S(\gamma_N(E_0))=g(E_0)$.
\smallskip

We see that $S(\gamma_{H(\rho_0,E_0,E)}(E))<S(\gamma_N(E))$ for all $E\neq E_0$ according to the optimality of $H(\rho_0,E_0,E)$.
Note also that the difference between the right and left sides of the last inequality is essential for $E\gg E_0$. The last fact is important for
the applications described in Section 6.
\end{example}

\section{Modified version of Theorem \ref{main}}

There is an important application (see Section 7.2)  which requires to modify the task solved by Theorem \ref{main}.
This modification consists in removing the condition (\ref{star}) from a positive operator $H$ in our optimization problem (leaving the condition (\ref{H-cond-w}) in force).

For a given state $\rho\in\S(\H)$ let $\M^+(\rho,E_0)$ be the set of all $\mathfrak{H}$-operators $H$ on $\H$ (defined in Section 2) satisfying  condition (\ref{H-cond-w}) such that $\Tr H\rho\leq E_{0}$, $E_0>0$. For given $E>0$ let $\M^+_E(\rho,E_0)$ be the subset of $\M(\rho,E_0)$ consisting of $\mathfrak{H}$-operators $H$ for each of which there exists the Gibbs state $\gamma_{H}(E)$ corresponding to the "energy" $E$ (see Section 2). It is clear that  the sets $\M^+(\rho,E_0)$ and $\M^+(\rho,E_0)$ substantially greater than  the sets $\M(\rho,E_0)$ and $\M(\rho,E_0)$ defined before Theorem \ref{main}.\smallskip

\begin{theorem}\label{main+-+}\emph{Let $\H$ be a separable infinite-dimensional Hilbert space and $\rho$ be a mixed state in $\S(\H)$ with the spectral representation $\rho=\sum\limits_{i=1}^{n} p_{i}|\varphi_i\rangle\langle \varphi_i|$ (where $n\leq +\infty$, $p_{i+1}\leq p_i$ and $p_i>0$ for all $\,i$) such that $S(\rho)<+\infty$. Let $\,E_0,E>0$ and $\,\theta=E/E_0$.} \smallskip

\emph{If $\theta\geq1$ then
\begin{equation*}
\inf_{H\in\M^+_E(\rho,E_0)}S(\gamma_{H}(E))=\inf_{H\in\M^+(\rho,E_0)}\sup_{\Tr H\sigma\leq E}S(\sigma)=S(\gamma_{H(\rho,E_0,E)}(E)),
\end{equation*}
where $H(\rho,E_0,E)$ is the operator described in Theorem \ref{main}.}
\smallskip

\emph{If $\theta<1$ then there exists an operator $H_+(\rho,E_0,E)$ in $\M^+_E(\rho,E_0)$ such that
\begin{equation}
S(\gamma_{H_+(\rho,E_0,E)}(E))=0.
\end{equation}
In this case}
$$
H_+(\rho,E_0,E)=E|\varphi_1\rangle\langle \varphi_1|+\frac{E_0-Ep_1}{1-p_1}\sum_{i=2}^{n}|\varphi_i\rangle\langle \varphi_i|\quad\textup{and}\quad \gamma_{H_+(\rho,E_0,E)}(E)=|\varphi_1\rangle\langle \varphi_1|.
$$
\end{theorem}

\emph{Proof.} Let $\theta\geq1$ and $H$ be an arbitrary operator in $\M^+(\rho,E_0)$.
Then
$$
\Tr H\langle\theta\rho\rangle_{\!\rm q}\leq\theta\Tr H\rho\leq\theta E_0=E,
$$
since the condition $\theta\geq1$ implies that $\langle\theta\rho\rangle_{\!\rm q}\leq\theta\rho$. Thus, Theorem \ref{main} shows that
$$
F_{H(\rho,E_0,E)}(E)=S(\langle\theta\rho\rangle_{\!\rm q})\leq F_{H}(E),
$$
where the inequality follows from the definition (\ref{F-def}) of the function $F_H$. Since $H$ is an arbitrary operator in $\M^+(\rho,E_0)$,
this inequality  implies the optimality of the operator $H(\rho,E_0,E)\in\M^+_E(\rho,E_0)$ among all the operators in $\M^+(\rho,E_0)$.\smallskip

The claim concerning the case $\theta<1$ is proved by noting
that
\begin{itemize}
  \item $\Tr H_+(\rho,E_0,E)\rho=E_0$;
  \item there is only one state $\sigma$ satisfying the condition
$\Tr H_+(\rho,E_0,E)\sigma\leq E$ -- the state $|\varphi_1\rangle\langle \varphi_1|$ (because $E<E_0$ implies $\frac{E_0-Ep_1}{1-p_1}>E$).
\end{itemize}
$\Box$

\section{Applications}

\subsection{General remarks}

The task described and solved in Section 3 was initially motivated by the following observation.

For many important characteristics of
$n$-partite infinite-dimensional quantum system $A_1...A_n$ (represented as a function $f$ on the set $\S(\H_{A_1..A_n})$) lower semicontinuity bounds of the form
\begin{equation}\label{sb-form}
f(\rho)-f(\sigma)\leq C\varepsilon F_H\!\left(\frac{\Tr H\rho_{A_1}}{\varepsilon}\right)+Dh^{\uparrow}(\varepsilon)
\end{equation}
were obtained (the list of these characteristics can be found below). Here,
\begin{itemize}
  \item $\rho$ and $\sigma$ are states in $\S(\H_{A_1..A_n})$ such that $d(\rho,\sigma)\leq\varepsilon$, where $d(\varrho,\varsigma)$ is a faithful measure of divergence between quantum states $\varrho$ and $\varsigma$;
  \item $H$ is a positive
operator satisfying conditions (\ref{H-cond-w}) and (\ref{star}) s.t. $\Tr H\rho_{A_1}<+\infty$;
  \item $C,D$ are the parameters depending on $f$;
  \item $F_H$ is the function defined in (\ref{F-def});
  \item  $h^{\uparrow}$ is the non-decreasing envelope of
the binary entropy defined as
\begin{equation}\label{h+}
h^{\uparrow}(p)=\left\{\begin{array}{l}
        h(p)\;\;\; \textrm{if}\;\;  p\in\shs[0,\frac{1}{2}]\\
        \ln2\;\quad \textrm{if}\;\;  p\in(\frac{1}{2},1]
        \end{array}\right., \quad h(p)\,\textrm{ is the binary entropy.}
\end{equation}
\end{itemize}
It is easy to show that for any state $\rho$ with finite  $S(\rho_{A_1})$  one can find (in uncountable number of ways)
a positive operator $H$ on $\H_{A_1}$ satisfying conditions (\ref{H-cond-w}) and (\ref{star}) such that $\Tr H\rho_{A_1}<+\infty$ and vice versa: the existence of
such an operator implies that $S(\rho_{A_1})<+\infty$ \cite[Propositions 1 and 4]{EC}. Naturally, the question arises \emph{how for given state $\rho$ and $\,\varepsilon>0$ to find
a positive operator $H$ on $\H_{A_1}$ satisfying conditions (\ref{H-cond-w}) and (\ref{star}) with finite $\Tr H\rho_{A_1}$ in such a way that the term
$$
F_H\!\left(\frac{\Tr H\rho_{A_1}}{\varepsilon}\right)
$$
be as small as possible.} The results of Section 3 give a direct answer to this question. Indeed, Theorem \ref{main} imply that
\begin{equation}\label{l-inf}
 \inf\left.\left\{F_H\!\left(\frac{\Tr H\rho_{A_1}}{\varepsilon}\right)\,\right|\, H\in \M(\rho_{A_1})\right\}=\textstyle S\!\left(H\!\left(\rho_{A_1},1,\frac{1}{\varepsilon}\right)\right)\doteq\displaystyle S\!\left(\left\langle\frac{\rho_A}{\varepsilon}\right\rangle_{\!\rm q}\right),
\end{equation}
where $\M(\rho_{A_1})$ is the set of positive operators $H$ on $\H_{A_1}$ satisfying conditions (\ref{H-cond-w}) and (\ref{star}) such that $\Tr H\rho_{A_1}<+\infty$,
$\,\langle\cdot\rangle_{\rm q}$ is the map defined in (\ref{psi-q}) and $\,S\,$ is the von Neumann entropy. Thus, we obtain the following\smallskip

\begin{proposition}\label{NEW-SCB} \emph{Let $f$ be a function on the set $\,\S(\H_{A_1..A_n})$  for which semicontinuity bound (\ref{sb-form}) holds and $\rho$ be a state in  $\S(\H_{A_1..A_n})$ such that $S(\rho_{A_1})<+\infty$. Then
\begin{equation}\label{NEW-SCB+}
f(\rho)-f(\sigma)\leq C\varepsilon S\!\left(\left\langle\frac{\rho_A}{\varepsilon}\right\rangle_{\!\rm q}\right)+Dh^{\uparrow}(\varepsilon)
\end{equation}for any state  $\sigma$ in $\S(\H_{A_1..A_n})$ such that $\,d(\rho,\sigma)\leq\varepsilon$.}
\end{proposition}\medskip

Corollary \ref{main-c} in Section 4 implies that the semicontinuity bound (\ref{NEW-SCB+}) is \emph{faithful}: the r.h.s. of (\ref{NEW-SCB+}) tends to zero as $\,\varepsilon\to0$ for
any state $\rho$ in $\S(\H_{A_1..A_n})$ with finite $S(\rho_{A_1})$.\medskip

As far as I know, at the moment\footnote{I would be grateful for any comments concerning this point.} the semicontinuity bounds of the form
(\ref{sb-form}) are obtained for the following characteristics\footnote{In the formulations of all the semicontinuity bounds mentioned below it is assumed that $H$ is a positive operator
satisfying the condition (\ref{H-cond}), which is stronger than condition (\ref{H-cond-w}). This assumption is necessary to guarantee
the faithfulness of  these semicontinuity bounds. It is easy to see (by using the proofs) that all the semicontinuity bounds are valid (as formal inequalities)
provided that the positive operator $H$ satisfies condition (\ref{H-cond-w}).}
\begin{itemize}
  \item the von Neumann entropy ($n=1$, $C=D=1$, $d(\varrho,\varsigma)=\frac{1}{2}\|\varrho-\varsigma\|_1$, Theorem 1 in \cite{BDJSH}).\footnote{The semicontinuity bound
(\ref{sb-form}) with $f=S$ is a simple corollary of the semicontinuity bound presented in Theorem 1 in \cite{BDJSH}.}
  \item the quantum conditional entropy of quantum-classical states ($n=2$, $C=D=1$, $d(\varrho,\varsigma)=\frac{1}{2}\|\varrho-\varsigma\|_1$, \cite[Proposition 7]{LCB+}).
  \item the entanglement of formation ($n=2$, $C=D=1$, $d(\varrho,\varsigma)=\sqrt{1-F(\varrho,\varsigma)}$, where $F(\varrho,\varsigma)=\|\sqrt{\varrho}\sqrt{\varsigma}\|_1^2$ is the fidelity of $\varrho$ and $\varsigma$, \cite[Proposition 10]{LCB+}).
  \item the quantum mutual information for commuting states ($n=2$, $C=D=2$, $d(\varrho,\varsigma)=\frac{1}{2}\|\varrho-\varsigma\|_1$, \cite[Example 1]{LCB+}).
  \item the quantum conditional mutual information for commuting states ($n=3$, $C=D=2$, $d(\varrho,\varsigma)=\frac{1}{2}\|\varrho-\varsigma\|_1$, \cite[Example 1]{LCB+}).
  \item the relative entropy of entanglement and the similar characteristics (the relative entropy of NPT entanglement, the Rains bound) for commuting states, ($n=2$, $C=D=1$, $d(\varrho,\varsigma)=\frac{1}{2}\|\varrho-\varsigma\|_1$, the remark at the end of Section 3 in \cite{LCB+}).
  \item the quantum discord with measured system $A_1$ for commuting states ($n=2$, $C=1$, $D=2$, $d(\varrho,\varsigma)=\frac{1}{2}\|\varrho-\varsigma\|_1$, the remark at the end of Section 3 in \cite{LCB+}).
  \item the quantum discord with measured system $A_2$ for commuting states ($n=2$, $C=D=2$, $d(\varrho,\varsigma)=\frac{1}{2}\|\varrho-\varsigma\|_1$, the remark at the end of Section 3 in \cite{LCB+}).
  \item the one-way classical correlation  with measured system $A_1$ for commuting states ($n=2$, $C=1$, $D=2$, $d(\varrho,\varsigma)=\frac{1}{2}\|\varrho-\varsigma\|_1$, the remark at the end of Section 3 in \cite{LCB+}).
  \item the one-way classical correlation with measured system $A_2$ for commuting states ($n=2$, $C=1$, $D=2$, $d(\varrho,\varsigma)=\frac{1}{2}\|\varrho-\varsigma\|_1$, the remark at the end of Section 3 in \cite{LCB+}).
\end{itemize}

For each of the above characteristics one can obtain an "optimized" semicontinuity bound using Proposition \ref{NEW-SCB}. By construction, this
 semicontinuity bound will be sharper than the "original" semicontinuity bound of the form (\ref{sb-form}). A concrete example can be found in the next subsection.\medskip

\subsection{New semicontinuity bound for the von Neumann entropy}

As mentioned in Section 7.1 Theorem 1 in \cite{BDJSH} implies the semicontinuity bound of the form
(\ref{sb-form}) for the von Neumann entropy. Namely, if $\rho$ is an arbitrary state in $\S(\H)$ and $H$ is  a positive
operator satisfying conditions (\ref{H-cond-w}) and (\ref{star}) s.t. $\Tr H\rho<+\infty$ then
\begin{equation}\label{SCB}
S(\rho)-S(\sigma)\leq \varepsilon F_H\!\left(\frac{\Tr H\rho}{\varepsilon}\right)+h^{\uparrow}(\varepsilon)
\end{equation}
for any state $\sigma$ in $\S(\H)$ such that $\frac{1}{2}\|\rho-\sigma\|_1\leq\varepsilon$.

Thus, Proposition \ref{NEW-SCB} implies the following "optimization" of the semicontinuity bound (\ref{SCB}): the inequality
\begin{equation}\label{SCB+}
S(\rho)-S(\sigma)\leq \varepsilon S\!\left(\left\langle\frac{\rho}{\varepsilon}\right\rangle_{\!\rm q}\right)+h^{\uparrow}(\varepsilon)
\end{equation}
holds for any states $\rho$ and $\sigma$ in $\S(\H)$ such that $\,\frac{1}{2}\|\rho-\sigma\|_1\leq\varepsilon$.\medskip

Inequality (\ref{SCB+}) is quite accurate but not optimal, because it is obtained from inequality (\ref{SCB}) which is slightly weaker than the optimal
semicontinuity bound for the von Neumann entropy presented in Theorem 1 in \cite{BDJSH}.

Using the method from the proof of Theorem 1 in \cite{FCB} and Theorem \ref{main+-+} in Section 6 one can improve inequality (\ref{SCB+}). For this purpose introduce the following notation.

For any positive trace class operator $\varsigma$ of rank $>1$ with  the spectral representation $\varsigma=\sum_{i=1}^{n} s_{i}|\phi_i\rangle\langle \phi_i|$ (where $n\leq +\infty$, $s_{i+1}\leq s_i$ and $s_i>0$ for all $\,i$) define positive trace class operator $\varsigma^{\dag}$ by the rule
\begin{equation}\label{d-d}
 \varsigma^{\dag}=\sum_{i=2}^{n} s_{i}|\phi_i\rangle\langle \phi_i|,
\end{equation}
i.e. $\varsigma^{\dag}$ is the operator obtained from $\varsigma$ by
removing the $1$-rank component corresponding to its maximal eigenvalue.

Now we can formulate the main result of this subsection.\smallskip

\begin{proposition}\label{SCB++} \emph{Let $\rho$ be a state in $\S(\H)$ such that $S(\rho)<+\infty$, $\sigma$ an arbitrary state in $\S(\H)$ and $\delta\doteq\frac{1}{2}\|\rho-\sigma\|_1$. Then
\begin{equation}\label{SCB+++}
   S(\rho)-S(\sigma)\leq \varepsilon S\!\left(\left\langle\frac{\rho^{\dag}}{\varepsilon}\right\rangle_{\!\rm q}\right)+h(\varepsilon),\qquad \varepsilon\doteq\min\{\delta,1-\|\rho\|\},
\end{equation}
where $(\cdot)^\dag$ and  $\,\langle\cdot\rangle_{\rm q}$ are the maps defined in (\ref{d-d}) and (\ref{psi-q}), $\|\rho\|$ is the operator norm of $\rho$, $h$ is the binary entropy. If $\,\delta<1-\|\rho\|$ then (\ref{SCB+++}) holds for any $\,\varepsilon\in[\delta,1-\|\rho\|]$.}\medskip

\emph{The inequality  (\ref{SCB+++}) is optimal: for any $\varepsilon\in[0,1]$ there exists states $\rho_\varepsilon$ and $\sigma_\varepsilon$ such that $"="$
holds in (\ref{SCB+++}). Moreover, the inequality (\ref{SCB+++}) implies
\begin{itemize}
  \item the optimal semicontinuity bounds for the von Neumann entropy with the rank constraint (presented in \cite[Proposition 2]{LCB}) which states that
\begin{equation}\label{S-CB-A}
   S(\rho)-S(\sigma)\leq \left\{\begin{array}{ll}
        \varepsilon\ln(r-1)+h(\varepsilon)&\textrm{if}\;\;  \varepsilon\leq 1-1/r\; \\
        \ln r& \textrm{if}\;\; \varepsilon\geq 1-1/r
        \end{array}\right.\!\!,\qquad  r=\rank\rho,
\end{equation}
for any finite rank state $\rho$ and any state $\sigma$ such that  $\,\frac{1}{2}\|\rho-\sigma\|_1\leq\varepsilon\in[0,1]$;
  \item the optimal semicontinuity bounds for the von Neumann entropy with the energy constraint (presented in \cite[Theorem 1]{BDJSH}) which states that
\begin{equation}\label{S-CB-B}
 S(\rho)-S(\sigma)\leq\left\{\begin{array}{l}
       \!\varepsilon F_{H_+}\!\!\left(\frac{E}{\varepsilon}\right)+h(\varepsilon)\quad\;\;\;\, \textrm{if}\;\;  \varepsilon\in[0,a_{H}(E)]\\\\
        \!F_{H}(E)\qquad \qquad\qquad\textrm{if}\;\;  \varepsilon\in[a_{H}(E),1]
        \end{array}\right.
\end{equation}
for any positive operator  $H$ satisfying conditions (\ref{H-cond}) and (\ref{star}), any state $\rho$ with $\Tr H\rho\leq E$ and any state $\sigma$ such that  $\,\frac{1}{2}\|\rho-\sigma\|_1\leq\varepsilon\in[0,1]$. Here,
\begin{equation}\label{H-rep+}
H_+\doteq \sum_{i=1}^{+\infty} h_{i+1}|\tau_i\rangle\langle \tau_i|,
\end{equation}
where it is assumed that $H$ has representation (\ref{H-form}), $a_{H}(E)=1-1/Z_{H}(E)$,  $F_{H}$, $F_{H_+}$  and $Z_H$ are  the functions defined in (\ref{F-def})  and (\ref{Z}).
\end{itemize}}
\end{proposition}\medskip

\emph{Proof.} Assume that $\sigma$ is a state such that $\,\frac{1}{2}\|\rho-\sigma\|_1\leq\varepsilon$. Let
$$
\rho=\sum_{i=1}^{+\infty} p_{i}|\varphi_i\rangle\langle \varphi_i|\quad\textup{and}\quad \sigma=\sum_{i=1}^{+\infty} q_{i}|\psi_i\rangle\langle \psi_i|
$$
be the spectral representations of $\rho$ and $\sigma$ such that $p_{i}\geq p_{i+1}$ and $q_{i}\geq q_{i+1}$ for all $i$.

Mirsky’s inequality (\ref{Mirsky-ineq+}) implies that
\begin{equation*}
     \frac{1}{2} ||\rho-\hat{\sigma}||_1 \leq\frac{1}{2} ||\rho-\sigma||_1\leq \varepsilon,
\end{equation*}
where
$$
\hat{\sigma}=\sum_{i=1}^{+\infty} q_{i}|\varphi_i\rangle\langle \varphi_i|
$$
Using Lemma 2A in \cite{FCB} it is easy to construct a state
$$
\sigma_*=\sum_{i=1}^{+\infty} q^*_{i}|\varphi_i\rangle\langle \varphi_i|
$$
such that
\begin{equation}\label{3-c}
q^*_1\geq p_1,\quad   S(\sigma_*)\leq S(\sigma)\quad\textup{and}\quad  \frac{1}{2} ||\rho-\sigma_*||_1\leq\frac{1}{2}||\rho-\hat{\sigma}||_1 \leq \varepsilon.
\end{equation}

Lemma \ref{sl} below and the first inequality in (\ref{3-c}) show that
\begin{equation}\label{u-in}
  \delta\doteq\frac{1}{2} ||\rho-\sigma_*||_1=\sum_{i=1}^{+\infty} |p_i-q^*_{i}|\leq c_\rho\doteq\sum_{i=2}^{+\infty} p_i=\Tr\rho^{\dag}.
\end{equation}

The fundamental entropic inequality (obtained in \cite{D++}) applied to the states $\rho$ and $\sigma_*$ implies
\begin{equation}\label{fin}
S(\rho)-S(\sigma_*)\leq \delta S(\tau_+)+h(\delta),
\end{equation}
where
$$
\tau_{+}=\frac{1}{\delta}[\rho-\sigma_*]_{+}=\frac{1}{\delta}\sum_{i=1}^{+\infty} [p_i-q^*_{i}]_{+}|\varphi_i\rangle\langle \varphi_i|\qquad ([x]_+=\max{0,x})
$$
is a state in $\S(\H)$.

Let $H_{\delta}\doteq H\!\left(\frac{\rho^{\dag}}{c_\rho},1,\frac{c_\rho}{\delta}\right)$ -- the "optimal" operator defined in Theorem 1 for the state $\frac{\rho^{\dag}}{c_\rho}$,
$E_0=1$ and $E=\frac{c_\rho}{\delta}$. The first inequality in (\ref{3-c}) shows that
$[p_1-q^*_{1}]_{+}=0$. It follows that $\delta\tau_{+}\leq \rho^{\dag}$ because
$\delta\tau_{+}\leq \rho$ by the construction. So,
$$
\Tr H_{\delta} \tau_{+}\leq \frac{1}{\delta}\,\Tr H_{\delta}\rho^{\dag}=\frac{c_\rho}{\delta}.
$$
Hence,
$$
S(\tau_{+})\leq F_{H_{\delta}}\!\left(\frac{c_\rho}{\delta}\right)=S\left(\left\langle\frac{\rho^{\dag}}{\delta}\right\rangle_{\!\rm q}\right),
$$
where the last inequality follows from the definition of $H\!\left(\frac{\rho^{\dag}}{c_\rho},1,\frac{c_\rho}{\delta}\right)$
and Theorem 1 in Section 3. This inequality, the second inequality in (\ref{3-c}) and (\ref{fin}) show that
\begin{equation}\label{fin+}
S(\rho)-S(\sigma)\leq S(\rho)-S(\sigma_*)\leq \delta S\left(\left\langle\frac{\rho^{\dag}}{\delta}\right\rangle_{\!\rm q}\right)+h(\delta).
\end{equation}

It follows from (\ref{3-c}) and (\ref{u-in}) that $\delta\leq \min\{\varepsilon, c_\rho\}$. Thus, to complete
the proof of (\ref{SCB+++}) it suffices to show that the function
$$
f(x)=x S\left(\left\langle\frac{c_\rho\hat{\rho}}{x}\right\rangle_{\!\rm q}\right)+h(x), \quad \hat{\rho}=\frac{\rho^{\dag}}{c_\rho},
$$
is not decreasing on $(0,c_\rho]$. Theorem \ref{main-p} in Section 4
shows that the function $G(E)=S\!\left(\left\langle E\hat{\rho}\right\rangle_{\!\rm q}\right)$ is
continuously differentiable and concave on $(0,+\infty)$. Moreover, it has the non-positive second derivative $G''(E)$ everywhere
on $(0,+\infty)$ excluding the points $E_2,E_3,...$ (belonging to $[1+\infty)$ and depending on the spectrum of $\hat{\rho}$). Hence the
function $f(x)$ is continuously differentiable on $(0,c_\rho]$ and has the  second derivative
$$
f''(x)=\left[G\left(\frac{c_\rho}{x}\right)-\frac{c_\rho}{x}G'\left(\frac{c_\rho}{x}\right)+\ln\frac{1-x}{x}\right]'_x=
\frac{c_\rho^2}{x^3}\,G''\left(\frac{c_\rho}{x}\right)-\frac{1}{x(1-x)}\leq 0
$$
for all  $x\in(0,c_\rho]$ excluding the points $\frac{c_\rho}{E_2}, \frac{c_\rho}{E_3},..$. It follows that the function $f(x)$ is concave $(0,c_\rho]$.
Thus, to prove that the function $f(x)$ is non-decreasing on $(0,c_\rho]$ we have to show that $f'(c_\rho)\geq 0$.

Note that
\begin{equation}\label{fd}
f'(c_\rho)=G(1)-G'(1)+\ln\frac{1-c_\rho}{c_\rho}.
\end{equation}
Theorem \ref{main-p} in Section 4 implies that $G(1)=S(\hat{\rho})$ and that
$G'(1)=S'_1(1)$,
where
$$
S_1(E)=Es_1+\eta(1-d_1E)+d_1\eta(E),\quad d_1=1-\frac{p_2}{c_\rho},\;\; s_1=S(\hat{\rho})-\eta\left(\frac{p_2}{c_\rho}\right).
$$
We have $S'_1(E)=s_1+d_1\ln\left(\frac{1}{E}-d_1\right)$ and, hence,
$$
G'(1)=S'_1(1)=s_1+d_1\ln(1-d_1).
$$
Substitution of $G(1)$ and $G'(1)$ into (\ref{fd}) gives
\begin{equation}\label{fd+}
\begin{array}{c}
f'(c_\rho)=S(\hat{\rho})-s_1-d_1\ln(1-d_1)+\ln\frac{1-c_\rho}{c_\rho}
=\eta\left(\frac{p_2}{c_\rho}\right)-d_1\ln\left(\frac{p_2}{c_\rho}\right)+\ln\frac{1-c_\rho}{c_\rho}\\\\
=-\ln\left(\frac{p_2}{c_\rho}\right)+\ln\frac{1-c_\rho}{c_\rho}=\ln\left(\frac{p_1}{p_2}\right)\geq 0,
\end{array}
\end{equation}
because $1-c_\rho=p_1$ and $p_1\geq p_2$.  Inequality (\ref{SCB+++}) is proved.\smallskip

To show the optimality of inequality (\ref{SCB+++}) it suffices to prove
that this inequality implies the optimal semicontinuity bound (\ref{S-CB-B}).

Let $H$ be a positive operator satisfying conditions (\ref{H-cond}) and (\ref{star}).  Let $\rho$ be a state such that $\Tr H\rho\leq E$ and $\sigma$ be a state such that $\,\frac{1}{2}\|\rho-\sigma\|_1\leq\varepsilon\in[0,1]$. W.l.o.g. we may assume that the operator
$H$ has representation (\ref{H-form}) in which $\{\tau_i\}_{i=1}^{+\infty}$ is the orthonormal system of eigenvectors of the state $\rho$ (arranged in such a way that
the corresponding sequence of eigenvalues is non-increasing).\footnote{If $\rho$ is a finite-rank  state then $\{\tau_i\}_{i=1}^{+\infty}$ is any extension of the
system of eigenvectors of $\rho$ to a basis in $\H$.} We may also assume that the operator
$H_+$ has the representation\footnote{We use the convention about the domain of an $\mathfrak{H}$-operator placed after (\ref{H-form}).} 
\begin{equation*}
H_+=\sum_{i=2}^{+\infty} h_{i}|\tau_i\rangle\langle \tau_i|.
\end{equation*}
The possibility of these assumptions follows from the Courant-Fischer theorem (cf.~\cite[Proposition II-3]{BDJ}) and from the fact that the functions $F_H$, $F_{H_+}$ and $a_H$ depend only on the spectra of the operators $H$ and $H_+$.

It follows from the proof of Theorem 1 in \cite{BDJSH} that for any $\varepsilon\in(0,1]$ the r.h.s. of (\ref{S-CB-B}) is equal to
\begin{equation*}
\max_{x\in[0,\shs\min\{\varepsilon,E/h_{2}\}]}\left(x F_{H_+}\!\left(\frac{E}{x}\right)+h(x)\right)
\end{equation*}
Thus, to show that  inequality (\ref{SCB+++}) implies the optimal semicontinuity bound (\ref{S-CB-B})
we have to prove that
\begin{equation}\label{t-in}
S\!\left(\left\langle\frac{\rho^{\dag}}{x}\right\rangle_{\!\rm q}\right)\leq F_{H_+}\!\left(\frac{E}{x}\right)\quad \forall x\in\left[0,c_\rho\right]
\end{equation}
(it is easy to see that the condition $\Tr H\rho\leq E$ implies $c_\rho\leq E/h_2$.)
 
Since $h_1=0$, we have $\Tr H_+\rho^\dag=\Tr H\rho\leq E$
and hence the operator $H_+$ belongs to the set $\M^+\!\left(\frac{\rho^{\dag}}{c_\rho},\frac{E}{c_\rho}\right)$ defined before Theorem \ref{main+-+} in Section 4.\footnote{
We cannot assert that this operator belongs to the set $\M\!\left(\frac{\rho^{\dag}}{c_\rho},\frac{E}{c_\rho}\right)$ defined before Theorem \ref{main} in Section 3, because
it may not satisfy condition (\ref{star}). So, here we have to use Theorem \ref{main+-+}  instead of Theorem \ref{main}.} So, $\frac{c_\rho}{E}H_+\in\M^+\!\left(\frac{\rho^{\dag}}{c_\rho},1\right)$.

Since $\frac{c_\rho}{x}\geq 1$, Theorem \ref{main+-+} implies that $H_x\doteq H\!\left(\frac{\rho^{\dag}}{c_\rho},1,\frac{c_\rho}{x}\right)$ is the optimal operator
in $\M^+\!\left(\frac{\rho^{\dag}}{c_\rho},1\right)$ and hence
$$
S\!\left(\left\langle\frac{\rho^{\dag}}{x}\right\rangle_{\!\rm q}\right)=S\!\left(\left\langle\frac{\rho^{\dag}}{c_\rho}\frac{c_\rho}{x}\right\rangle_{\!\rm q}\right)=F_{H_x}\!\left(\frac{c_\rho}{x}\right)\leq F_{\frac{c_\rho}{E}H_+}\!\left(\frac{c_\rho}{x}\right)=
F_{H_+}\!\left(\frac{E}{x}\right),$$
where the second equality follows from Theorem \ref{main}, the inequality -- from Theorem \ref{main+-+}, the last equality -- from definition 
(\ref{F-def}) of the function $F_H$.    

To show that the inequality (\ref{SCB+++}) implies the optimal semicontinuity bound (\ref{S-CB-A}) it suffices to note that
the r.h.s. of (\ref{S-CB-A}) is equal to
\begin{equation*}
\max_{x\in[0,\varepsilon]}\left( x \ln(r-1)+h(x)\right)
\end{equation*}
and that
$$
S\left(\left\langle\frac{\rho^{\dag}}{x}\right\rangle_{\!\rm q}\right)\leq \ln(r-1)
$$
because $\,\rank\rho^{\dag}=\rank\rho-1=r-1$ by the construction of $\rho^{\dag}$  $\Box$. \medskip

The following lemma is proved by simple arguments.\smallskip

\begin{lemma}\label{sl} \emph{If $\,\{p_i\}_{i=1}^{+\infty}$ and $\,\{q_i\}_{i=1}^{+\infty}$ are probability
distributions such that $p_1\leq q_1$ then}
$$
\sum_{i=1}^{+\infty}|p_i-q_i|\leq 2(1-p_1).
$$
\end{lemma}

\begin{remark}\label{dd}
The second claim of Proposition \ref{SCB++} shows that we may treat (\ref{SCB+++})  
as "generating inequality" for the optimal  semicontinuity bounds on the von Neumann entropy with the rank/energy constraint.  

In fact, inequality (\ref{SCB+++}) is stronger than both these semicontinuity bounds, which does not contradict to their optimality,
because it uses more information about the state $\rho$ than these bounds).\footnote{For a similar reason,  the state-dependent improvements of the Audenaert's
continuity bound for the von Neumann entropy obtained recently in \cite{D++,D+++} do not contradict to the optimality of this bound.}   
This is confirmed by numerical analysis of the accuracy of semicontinuity bound (\ref{SCB+++}) which will be presented
in the companion article \cite{ISB}.
\end{remark}

\subsection{New semicontinuity bound for the Entanglement of Formation}

The Entanglement of
Formation (EoF) is one of the basic entanglement measures in bipartite quantum systems \cite{Bennett,4H,P&V}.
If  infinite-dimensional bipartite quantum system $AB$  there are two versions $E_F^d$ and $E_F^c$ of the EoF defined, respectively, by means of discrete and continuous convex roof extensions
\begin{equation}\label{E_F-def-d}
E_F^d(\omega)=\!\inf_{\sum_k\!p_k\omega_k=\omega}\sum_kp_kS([\omega_k]_A),
\end{equation}
\begin{equation}\label{E_F-def-c}
E_F^c(\omega)=\!\inf_{\int\omega'\mu(d\omega')=\omega}\int\! S(\omega'_A)\mu(d\omega'),
\end{equation}
where $\omega$ is a state in $\S(\H_{AB})$,  the  infimum in (\ref{E_F-def-d}) is over all countable ensembles $\{p_k, \omega_k\}$ of pure states in $\S(\H_{AB})$ with the average state $\omega$ and the  infimum in (\ref{E_F-def-c}) is over all Borel probability measures on the set of pure states in $\S(\H_{AB})$ with the barycenter $\omega$ (the last infimum is always attained). It follows from the definitions that $E_F^d(\omega)\geq E_F^c(\omega)$, while it is known that $E_F^d(\omega)=E_F^c(\omega)$ for any state $\omega$ in $\S(\H_{AB})$ such that $\min\{S(\omega_{A}),S(\omega_{B})\}<+\infty$ \cite[Section 4.4]{QC}. The conjecture of coincidence of $E_F^d$ and $E_F^c$ on $\S(\H_{AB})$ is an interesting open question.

Proposition 10 in \cite{LCB+} and the remark after it imply that the semicontinuity bound (\ref{sb-form}) holds for both functions $E_F^d$ and $E_F^c$ with
$d(\varrho,\varsigma)=\sqrt{1-F(\varrho,\varsigma)}$, where $F(\varrho,\varsigma)=\|\sqrt{\varrho}\sqrt{\varsigma}\|_1^2$ is the fidelity of $\varrho$ and $\varsigma$.
So, Proposition \ref{NEW-SCB} in Section 6.1 implies the following\smallskip

\begin{proposition}\label{EF-SCB} \emph{Let $AB$ be an infinite-dimensional bipartite quantum system.}\smallskip

\emph{If $\,\rho$ is a state in $\S(\H_{AB})$ such that $S(\rho_A)<+\infty$ then
\begin{equation}\label{EF-SCB-B}
   E^*_F(\rho)-E^*_F(\sigma)\leq \varepsilon S\left(\left\langle\frac{\rho_A}{\varepsilon}\right\rangle_{\!\rm q}\right)+h^{\uparrow}(\delta),\quad E^*_F=E_F^d,E_F^c,
\end{equation}
for any state $\sigma$ in $\S(\H_{AB})$ such that $\,\sqrt{1-F(\rho,\sigma)}\leq \varepsilon\leq1$, where
$\,\langle\cdot\rangle_{\rm q}$ is the map defined in (\ref{psi-q}), $\,S\,$ is the von Neumann entropy and  $\,h^{\uparrow}$ is the function defined in (\ref{h+}).}
\end{proposition}\medskip

Corollary \ref{main-c} in Section 4 implies that the semicontinuity bound (\ref{EF-SCB-B}) is \emph{faithful}: the r.h.s. of (\ref{EF-SCB-B}) tends to zero as $\,\varepsilon\to0$ for
any state $\rho$ in $\S(\H_{AB})$ with finite $S(\rho_{A})$.\medskip

\textbf{Note:}  It is easy to show that the semicontinuity bound (\ref{EF-SCB-B}) implies  the  semicontinuity bounds with the rank/energy constraints
presented in parts A and B of Proposition 10 in \cite{LCB+}. In fact, the semicontinuity bound (\ref{EF-SCB-B}) significantly improves both of these bounds.\smallskip

By the well-known inequality
\begin{equation}\label{F-Tn-ineq}
1-\sqrt{F(\varrho,\varsigma)}\leq\textstyle\frac{1}{2}\|\varrho-\varsigma\|_1
\end{equation}
we may replace the condition $\,\sqrt{1-F(\varrho,\varsigma)}\leq \varepsilon\leq1$ in Proposition \ref{EF-SCB} with the
condition $\,\sqrt{\epsilon(2-\epsilon)}\leq \varepsilon\leq1$, where $\,\epsilon=\frac{1}{2}\|\varrho-\varsigma\|_1$. Of course, such replacement leads to a loss of accuracy of
the bound (\ref{EF-SCB-B}).\smallskip

A numerical analysis of the accuracy of semicontinuity bound (\ref{EF-SCB-B}) and other semicontinuity bounds obtained by using Proposition \ref{NEW-SCB}  will be presented
in the companion article \cite{ISB}.

\section*{Appendix}

\subsection*{A-1. Basic lemmas}

The proof  of Theorem \ref{main} is essentially based on the following Lemma \ref{bl}. In this lemma we
will use the maps $\mu_{\rm c}$ and $\langle\cdot\rangle_{\rm c}$  described at the beginning of Section 3.
\smallskip

\begin{lemma}\label{bl} \emph{Let $\{a_i\}_{i=1}^n$, $2\leq n\leq+\infty$, be an $n$-tuple of positive numbers arranged in the non-increasing order such that $\sum_{i=1}^na_i\leq1$. Let $I_n\doteq \N\cap[1,n].$\footnote{Here and in what follows we assume that $(a,b]=(a,+\infty)$ if $b=+\infty$.} \smallskip
Let $\,\theta>0$ and
$$
\Omega_0\doteq\left\{\{x_i\}_{i=1}^n\,\left|\, x_1=0,\; x_i\in \mathbb{R},\;\,x_{i}\leq x_{i+1}\;\, \forall i\in I_{n-1},\; \sum_{i=1}^na_ix_i=1\right.\right\}
$$
Write $A$ and $B$ for the following cases:
\begin{itemize}
 \item $\,n<+\infty\,$ and  $\,a_nn\geq 1/\theta \;$ (case $A$);
 \item either $\,n=+\infty\,$ or $\,n<+\infty\,$ and $\,a_nn< 1/\theta\;$  (case $B$).
\end{itemize}
Then
\!\!\begin{equation}\label{bl+}
\begin{array}{c}
 \displaystyle\inf\left\{\left. \theta b+\ln\left(\sum_{i=1}^ne^{-bx_i}\right)\,\right|\,b\geq 0,\,\{x_i\}_{i=1}^n\in\Omega_0\right\}=
\theta b_*+\ln\left(\sum_{i=1}^ne^{-b_*x^*_i}\right) \\\\
 \displaystyle =H\!\left(\langle(\theta a_1, \theta a_2, \theta a_3,...)\rangle_{\rm c}\right)\\\\= \left\{\begin{array}{ll}
        \ln n\;&\textit{in case } A \\
        \eta(1-\theta d_{m})+(1-\theta d_m)\ln m+\theta s_m+d_m\eta(\theta)\; & \textit{in case } B
        \end{array}\right.\!,
\end{array}
\end{equation}
where $H$ is the Shannon entropy, $\langle\cdot\rangle_{\rm c}$ is the map defined in (\ref{psi-c}),
\begin{eqnarray}
d_k&=&\sum_{i=k+1}^{n}a_i,\quad k\in\{0\}\cup I_{n-1},\label {d_k}\\
s_k&=&\sum_{i=k+1}^{n}\eta(a_i), \quad k\in\{0\}\cup I_{n-1}, \label {s_k}\\
m &=&\mu_{\rm c}((\theta a_1, \theta a_2, \theta a_3,...)),\\
\nonumber b_*&=&\left\{\begin{array}{ll}
       0&\textit{in case } A\medskip\\
       s_m+d_m\ln\frac{1-\theta d_m}{\theta m}\;&\textit{in case } B
       \end{array}\right., \\
\nonumber\{x^*_i\}_{i=1}^n&=&\left\{\begin{array}{ll}
       \textit{any }n\textit{-tuple in } \,\Omega_0 &\textit{in case } A\medskip\\
        (\underbrace{0,...,0}_{m\;\textit{items}},c_{m+1},c_{m+2},...),\;\;c_i=\frac{1\,}{b_*} \ln \frac{1-\theta d_m}{a_i\theta m},&\textit{in case } B
        \end{array}\right., \\\end{eqnarray}
$\eta(x)=-x\ln x$. In case B the "optimal" pair $\left(b_*,\{x^*_i\}_{i=1}^n\right)$ is unique and satisfies the condition
\begin{equation}\label{imp-eq}
  \sum_{i=1}^{n}x^{*}_ie^{-b_*x^{*}_i}=\theta\sum_{i=1}^{n}e^{-b_*x^{*}_i}.
\end{equation}}
\end{lemma}

\emph{Proof.} The main claims of the lemma follows from Lemma \ref{ml2} below.

To prove the second equality in (\ref{bl+}) it is necessary
to:
\begin{itemize}
  \item note that in case A we have $\,\langle(\theta a_1, \theta a_2, \theta a_3,...)\rangle_{\rm c}=(\underbrace{1/n,...,1/n}_{n\textrm{ items}})$;
  \item note that in case B we have $\,\langle(\theta a_1, \theta a_2, \theta a_3,...)\rangle_{\rm c}=(\underbrace{c/m,...,c/m}_{m\textrm{ items}}, \theta a_{m+1},\theta a_{m+2},...)$,
  where $\,c=1-\theta\sum_{i=m+1}^na_i$;
  \item use Lemma \ref{ml2} below to show that the l.h.s. of (\ref{bl+}) is equal to the last expression in (\ref{bl+}),
  \item use the identity $\,\eta(ab)=a\eta(b)+\eta(a)b\,$ to transform the case $B$ formula in the last expression in (\ref{bl+}) into the form
  $$
  \sum_{i=m+1}^n \eta(\theta a_i)+\eta(c)+c\ln m=H\left((c,\theta a_{m+1},\theta a_{m+2},...)\right)+c\ln m;
  $$
  \item note that
  $$
  \begin{array}{c}
  H(c,\theta a_{m+1},\theta a_{m+2},...)+c\ln m=\bar{c} H\!\left((\theta/\bar{c}) a_{m+1},(\theta/\bar{c}) a_{m+2},...)\right)+h(c)+c\ln m\\\\=H\!\left((\underbrace{c/m,...,c/m}_{m\textrm{ items}}, \theta a_{m+1},\theta a_{m+2},...)\right),
  \end{array}
  $$
  where $\,\bar{c}=1-c$ and $\,h$ is the binary entropy.
 \end{itemize}

To prove (\ref{imp-eq}) it suffices to use the  expressions for $b_*$ and $\,x_i^{*}$ and to note that
$$
\ln \frac{1-\theta d_m}{\theta m}=\frac{b_*-s_m}{d_m}
$$
Indeed, direct calculations show that both sides of (\ref{imp-eq}) are equal to
$\frac{\theta m}{1-\theta d_m}$. $\Box$ \medskip

\begin{lemma}\label{ml2} \emph{Let the conditions and notation of Lemma \ref{bl} be valid and} \footnote{It is clear that $\,z_0(0)=n\leq+\infty$.}
\begin{equation}\label{z-d}
z_0(b)\doteq\inf\left\{\left.\sum_{i=1}^ne^{-bx_i}\,\right|\, \{x_i\}_{i=1}^n\in\Omega_0\right\},\quad b\geq0.
\end{equation}

\emph{For any $\,b>0$ the infimum in (\ref{z-d}) is attained at a unique  $n$-tuple $\{x^b_i\}_{i=1}^n\in\Omega_0$ and}\footnote{The r.h.s. of (\ref{Z-def+}) is well defined for
for any $\,b>0,$  see Remark \ref{ml-r} below.}\medskip
\begin{equation}\label{Z-def+}
z_0(b)=\sum_{k=1}^{n-1}\mathbf{1}_{[b_{k+1},b_{k})}(b)\left( k+d_ke^{\frac{s_k-b}{d_k}}\right),
\end{equation}
where
\begin{equation}\label{b-new}
\!b_1=+\infty,\quad b_k=s_{k-1}+d_{k-1}\ln a_k,\quad k\in I_{n-1}\setminus\{1\}\quad \textit{and} \quad b_n=0\;\;\; \textit{if} \;\;\; n<+\infty.
\end{equation}

\emph{If $\,n<+\infty\,$ then for any $\,\theta>0$ the function $f(b)=\theta b+\ln z_0(b)$ on $[0,+\infty)$ has the following properties:
\begin{itemize}
  \item the function $f(b)$ is positive, convex and continuously differentiable on $[0,+\infty)$;~\footnote{One can show that the second derivative $f''(b)$ is not continuous.}
  \item $f(0)=\ln n\,$, $f'(0+)=\theta-\frac{1}{a_nn}\,$ and
$\;\,\lim_{b\to+\infty}f(b)=+\infty$; \footnote{$f'(0+)$ denotes the right derivative of $f$  at the point $0$.}
  \item
\begin{equation}\label{lr-2}
        \inf_{b\in[0,+\infty)}f(b)=f(b_*),
\end{equation}
where\footnote{It is easy to see that the $(n-1)\textup{-}$tuple $\{c_k\doteq d_k+ka_k\}^{n-1}_{k=1}$ is non-increasing and that $c_1=d_0$. So, $m$ is
the maximal $k$ such that $\,c_k> 1/\theta\,$ if $\,\theta> 1/d_0\,$ and $\, m=1\,$ if $\,\theta\leq1/d_0$.}
\begin{eqnarray}
  b_*&=&\left\{\begin{array}{ll}
        0&\textit{if}\;\; a_nn\geq 1/\theta\;\,  \quad (\textit{case } A)\medskip\\
        s_m+d_m\ln\frac{1-\theta d_m}{\theta m}\;&\textrm{if}\;\;  a_nn< 1/\theta \quad\;\, (\textit{case } B)
        \end{array}\right., \label{bst-d}\\
  m &=&\left\{\begin{array}{ll}
        1&\textit{if}\;\;   \theta<1/d_0  \medskip\\
        \max\{k\,|\,d_k+ka_k>1/\theta\}\;\;& \textit{if}\;\; \theta\geq 1/d_0
        \end{array}\right., \label{ell-d}\\
  \nonumber f(b_*)&=&\left\{\begin{array}{ll}
        \ln n&\textit{in case } A\medskip \\
        b_*\theta+\ln\frac{m}{1-\theta d_m } & \textit{in case } B\\
        \end{array}\right..
\end{eqnarray}
Moreover, $b_*$ is a unique point in
$[0,+\infty)$ at which the infimum of $f$ is attained.
\end{itemize}}

\emph{In  case $A$  the role of $\,\{x_i^{b_*}\}_{i=1}^n$ is played by any $n$-tuple in $\Omega_0$.}\smallskip

\emph{In  case $B$  the unique $n$-tuple $\{x_i^{b_*}\}_{i=1}^n$ consists of the elements
$$
        x_i^{b^*} = \left\{\begin{array}{ll}
        0&\textrm{if}\;\;  1\leq i\leq m \\
        \frac{1\,}{b_*} \ln \frac{1-\theta d_m}{a_i\theta m}& \textrm{if}\;\; m< i\leq n
        \end{array}\right.,
$$
where $b_*$ and $m$ are the parameters defined in (\ref{bst-d}) and (\ref{ell-d}).}\smallskip

\emph{If $\,n=+\infty$ then any $\theta>0$ the function $f(b)=\theta b+\ln z_0(b)$  on $(0,+\infty)$ has the following properties:
\begin{itemize}
  \item the function $f(b)$ is positive, convex and continuously differentiable on $(0,+\infty)$;
  \item
  \begin{equation}\label{lr}
    \lim_{b\to0^+}f(b)=+\infty, \quad\lim_{b\to0^+}f'(b)=-\infty\quad and \quad\lim_{b\to+\infty}f(b)=+\infty;
  \end{equation}
  \item  the relation (\ref{lr-2}) with $\,[0,+\infty)$ replaced by $\,(0,+\infty)$  holds in which
$$
b_*=s_m+d_m\ln\frac{1-\theta d_m}{\theta m},
$$
where $m$ is defined in (\ref{ell-d}), and
$$
\begin{array}{rl}
  f(b_*)\!&=\;b_*\theta+\ln\frac{m}{1-\theta d_m }\\\\
  &=\; \eta(1-\theta d_{m})+(1-d_m)\ln m+\theta s_m+d_m\eta(\theta),\quad (\eta(x)=-x\ln x).
\end{array}
$$
Moreover, $b_*$ is a unique point in
$(0,+\infty)$ at which the infimum of $f$ is attained.
\end{itemize}
The $n$-tuple $\{x_i^{b_*}\}_{i=1}^n$ consists of the elements}
$$
  x_i^{b_*}=\left\{\begin{array}{ll}
        0&\textrm{if}\;\;  1\leq i\leq m \\
        \frac{1\,}{b_*} \ln \frac{1-\theta d_m}{a_i\theta m}& \textrm{if}\;\; i>m
        \end{array}\right.
$$
\end{lemma}
\smallskip

\begin{remark}\label{ml-r} It is easy to see that
$b_2\geq b_3\geq b_4... $ and that $b_k\to0$ as $k\to+\infty$ in the case $\,n=+\infty$.
So, in the case $\,n<+\infty\,$ (resp.  $\,n=+\infty\,$)  $\,\bigcup_{k=1}^{n-1}[b_{k+1},b_k)$ coincides with $\,[0,+\infty)$ (resp. $\,(0,+\infty)$).

Some of the sets $[b_{k+1},b_k)$ may be empty, as $\,a_k=a_{k+1}\,$ for some $k>1$ implies  that $\,b_k=b_{k+1}$ and vice versa because
it is easy to see that
\begin{equation}\label{b-r-f}
b_{k+1}=b_{k}-d_{k}\ln \left(\frac{a_{k}}{a_{k+1}}\right)\quad  \forall k\in I_{n-1}\cap[2,+\infty).
\end{equation}
\end{remark}\smallskip

\emph{Proof.} Properties of the function $f(b)$ for any $\,n\leq+\infty\,$ are proved by direct calculation of the first and second derivatives of the function $f(b)$ (via calculation of the first and second derivatives of the function $z_0(b)$) with the use of Lemma \ref{ml} below.

To show that both functions $f(b)$ and $f'(b)$ are continuous
we have to show continuity of the functions $z_0(b)$ and $z'_0(b)$ at the "gluing points" $b_2,b_3,..$ keeping into account that some of these points may coincide.

Note that in the case $\,a_1=a_2=...=a_n$, $n<+\infty$, there are no "gluing points", since in this case $\,b_2=b_3=...=b_n=0$.

Assume that  $a_{k+1}=...=a_{k+m}>a_{k+m+1}$ for some $k\in\{0\}\cup I_{n-1}$ and $m\in\N$ and that $a_k>a_{k+1}$ in the case $k>0$. Then it follows from
(\ref{b-r-f}) that $b_p>b_{p+1}=...=b_{k+m}>b_{k+m+1}$, where $p=1$ in the case $k=0$ and $p=k$ otherwise. Then for all $b$ and $b'$ such that $b_{k+m+1}\leq b<b_{k+m}=...=b_{p+1}\leq b'<b_{p}$
we have
$$
z_0(b)=k+m+d_{k+m}\exp\left(\frac{s_{k+m}-b}{d_{k+m}}\right),\quad z_0(b')=p+d_{p}\exp\left(\frac{s_{p}-b'}{d_{p}}\right)
$$
and
$$
z'_0(b)=-\exp\left(\frac{s_{k+m}-b}{d_{k+m}}\right),\quad z'_0(b')=-\exp\left(\frac{s_{p}-b'}{d_{p}}\right).
$$
So, by using the definitions (\ref{d_k}), (\ref{s_k}) and (\ref{b-new}) it is easy to show that
\begin{equation}\label{d-1}
z_0(x_*-)=k+m+\frac{d_{k+m}}{a_{k+m}}=p+\frac{d_{p}}{a_{p+1}}=z_0(x_*+)
\end{equation}
and
\begin{equation}\label{d-2}
z'_0(x_*-)=-\frac{1}{a_{k+m}}=-\frac{1}{a_{p+1}}=z'_0(x_*+),
\end{equation}
where $\,x_*=b_{p+1}=...=b_{k+m}\,$ and $\,y(x-)\,$ (resp. $y(x+)$) denote the left (resp. right) limit of a function $y(x)$
at a point $x$. Note that formulae (\ref{d-1}) and (\ref{d-2}) imply that
\begin{equation}\label{d-3}
f(x_*)=\theta x_*+\ln \left( p+\frac{d_{p}}{a_{p+1}}\right)\quad\textrm{ and }\quad f'(x_*)=\theta-\frac{1}{pa_{p+1}+d_p}.
\end{equation}

Since the functions $f$ and $f'$ are continuous, to prove the convexity of $f$ on $\mathbb{R}_+$ it suffices to note that
$$
f''(b)=\frac{k\exp\left(\frac{b-s_{k}}{d_{k}}\right)}{d_k\left[k\exp\left(\frac{b-s_{k}}{d_{k}}\right)+d_k\right]^2}>0
$$
for any $\,b\in (b_{k+1},b_k)$, $k\in I_n$.\smallskip

If $\,n<+\infty\,$ then direct calculation shows that $\,f(0)=\ln n\,$ and
$\,f'(0+)=\theta-\frac{1}{a_nn}$. If $n=+\infty$ then the formulae in (\ref{d-3}) implies the first and the second limit relations in (\ref{lr}).
The third limit relations in (\ref{lr}) follows from the fact that
$$
f(b)=\theta b+\ln\left[1+d_1\exp\left(\frac{s_{1}-b}{d_{1}}\right)\right]\geq \theta b
$$
for any $\,n\leq+\infty\,$ and all $\,b\,$ large enough. \smallskip

Since the function $f$ is strictly convex and continuously differentiable, it achieves its infimum on $\mathbb{R}_+$ at the point
$$
b_*=\left\{\begin{array}{ll}
        0&\textup{if }\, n<+\infty\,\textrm{ and }\, f'(0)\geq 0\;  \quad (\textup{case } A)\medskip\\
        x_*\;&\textrm{if either }\,  n<+\infty\,\textrm{ and }\, f'(0)< 0\,\textrm{ or }\,n=+\infty \quad (\textup{case } B)
        \end{array}\right.,
$$
where $x_*$ is the unique solution of the equation $f'(x)=0$.

To determine $x_*$ in case $B$ note that
$$
f'(b)=\theta-\frac{1}{k\exp\left(\frac{b-s_{k}}{d_{k}}\right)+d_k}
$$
for any $\,b\in [b_{k+1},b_k)$, $k\in I_{n-1}$.  So,
$$
x_*=s_m+d_m\ln \left(\frac{1-d_m\theta}{\theta m}\right),
$$
where $m\in I_{n-1}$ is such that the r.h.s. of this equality belongs to $[b_{m+1},b_m)$. This number
$\,m\,$ is unique and always exists by the above arguments.

To find $m$ note that
$$
s_m+d_m\ln \left(\frac{1-d_m\theta}{\theta m}\right)=b_{m+1}+d_m\ln \left(\frac{1-d_m\theta}{\theta m a_{m+1}}\right).
$$
Thus, for $m>1$ the condition $x_*\in[b_{m+1},b_m)$ is equivalent to the condition
$$
0\leq d_m\ln \left(\frac{1-d_m\theta}{\theta ma_{m+1}}\right)< b_m-b_{m+1}=d_m\ln \left(\frac{a_m}{a_{m+1}}\right),
$$
where the last equality follows from (\ref{b-r-f}). So, in this case $m$ is determined by the inequalities
$$
c_{m+1}\leq \frac{1}{\theta}<c_m,
$$
where $\{c_k\doteq d_k+ka_k\}^{n-1}_{k=1}$ is a non-increasing $(n-1)\textup{-}$tuple such that $c_1=d_0$. So, $\,m\,$ is
the maximal $k$ such that $\,c_k> 1/\theta$, i.e. $\,m\,$ is determined by formula (\ref{ell-d}).

The condition
$\,x_*=s_1+d_1\ln \left(\frac{1-d_1\theta}{\theta}\right)\in[b_{2},+\infty)\,$
holds if and only if $\,\frac{1}{\theta}\geq c_2$. So, in this case $\,m=1\,$ is also determined by formula (\ref{ell-d}).\smallskip

The  expressions for $\,x_i^{b_*}$ in case $B$ can be established using Lemma \ref{ml} below (with $c=0$).
$\Box$

\medskip
\begin{lemma}\label{ml} \emph{Let $\{a_i\}_{i=1}^n$, $2\leq n\leq+\infty$, be an $n$-tuple of positive numbers arranged in the non-increasing order such that $\sum_{i=1}^na_i\leq1$. Let $I_n\doteq \N\cap[1,n]$.} \smallskip

\emph{For any $b>0$ and $c\geq0$ there exists a unique $n$-tuple  $\,\bar{x}_b=\{x^b_i\}_{i=1}^n$ of nonnegative numbers belonging to the set
$$
\Omega_{\rm c}\doteq\left\{\{x_i\}_{i=1}^n\,\left|\, x_1\in [0,c],\; x_i\in \mathbb{R},\;\,x_{i}\leq x_{i+1}\;\, \forall i\in I_{n-1},\; \sum_{i=1}^na_ix_i=1\right.\right\}
$$
such that
\begin{equation}\label{ml+}
\sum_{i=1}^ne^{-bx^b_i}=\inf\left\{\left.\sum_{i=1}^ne^{-bx_i}\,\right|\, \{x_i\}_{i=1}^n\in\Omega_{\rm c}\right\}.
\end{equation}
Let $\,\{b_k\}_{k=1}^n$ be a non-increasing $n$-tuple  of nonnegative numbers defined as
\begin{eqnarray}
    b_0 &=& \left\{\begin{array}{ll}
        \frac{s_0+d_0\ln a_1}{1-cd_0}&\textrm{if}\;\;   cd_0<1 \\
        +\infty & \textrm{if}\;\; cd_0\geq1 \end{array}\right.,
      \\
    b_k &=& s_{k-1}+d_{k-1}\ln a_k, \quad k\in I_{n}, \label{b-k-def}\\
    b_n &=& 0 \;\textrm{ if }\;n<+\infty,
\end{eqnarray}
where
\begin{equation}\label{d-s-def}
 d_k=\sum_{i=k+1}^{n}a_i,\quad
 s_k=\sum_{i=k+1}^{n}\eta(a_i), \quad k\in\{0\}\cup I_{n-1} \quad(\eta(x)=-x\ln x).
\end{equation}
If $\,b\in [b_{1},b_0]$ then
  $$
  x^b_i=\frac{1}{b}\left( \frac{b-s_0}{d_0} -\ln a_i\right),\;\; i\in I_n, \quad  \sum_{i=1}^ne^{-bx^b_i}=d_0e^{\frac{s_0-b}{d_0}}.
  $$
If $\,b\in [b_{k+1},b_k)\cap(0,+\infty)$, $\,k\in I_{n-1}$, then
$$
x^b_i=\left\{\begin{array}{ll}
        0&\textrm{if}\;\;  1\leq i\leq k \\
        \frac{1}{b}\left( \frac{b-s_k}{d_k} -\ln a_i\right)& \textrm{if}\;\;  i\in I_n\setminus\{1,..,k\}
       \end{array}\right.\!\!,\quad  \sum_{i=1}^{n}e^{-bx^b_i}=k+d_ke^{\frac{s_k-b}{d_k}}.
$$
If $\,b_{0}<+\infty$ and $\,b>b_0$ then
$$
x^b_i=\left\{\begin{array}{ll}
        c&\textrm{if}\;\;   i=1 \\
        \frac{1}{b}\left( \frac{b(1-a_1c)-s_1}{d_1} -\ln a_i\right)& \textrm{if}\;\; i\in I_n\setminus\{1\}
       \end{array}\right.\!\!\!, \;\;\sum_{i=1}^{n}e^{-bx^b_i}=e^{-bc}\!\left(1+d_1e^{\frac{s_1-b(1-cd_0)}{d_1}}\right).
$$}

\emph{The function $\,b\mapsto x^b_i\,$ is continuous on $\,(0,+\infty)\,$ for each $\,i\in I_n$.}\smallskip

\emph{For any $b>0$ and $c\geq0$ the r.h.s. of (\ref{ml+}) can be expressed as\footnote{$\mathbf{1}_{B_k}$ is the indicator function of the set $B_k$.}
\begin{equation}\label{Z-def+c}
z_{\rm c}(b)=\sum_{k=0}^{n-1}\mathbf{1}_{B_k}(b)\left( k+d_ke^{\frac{s_k-b}{d_k}}\right)+\mathbf{1}_{B_c}(b)\, e^{-bc}\!\left(1+d_1e^{\frac{s_1-b(1-cd_0)}{d_1}}\right),
\end{equation}
where $\,B_{k}=[b_{k+1},b_{k})$, $\;k\in \{0\}\cup I_{n-1}$, and $\,B_{c}=[b_{0},+\infty)$.}\smallskip

\emph{If $\,c=0\,$ then  $\,b_1=b_0$ and the r.h.s. of (\ref{Z-def+c}) coincides with the r.h.s. of (\ref{Z-def+}).}\medskip
\end{lemma}

\begin{remark}\label{ml-r+c} In the case $\,n<+\infty\,$ (resp.  $\,n=+\infty\,$) the union of $[b_0,+\infty)$ and $\bigcup_{k=0}^{n-1}[b_{k+1},b_k)$ coincides with $\,[0,+\infty)$ (resp. $\,(0,+\infty)$).  Indeed, if $\,n<+\infty\,$ then $b_n=0$, if $\,n=+\infty\,$ then the non-increasing sequence $\{b_k\}$ tends to $0$ as $k\to+\infty$.

Some of the sets $[b_{k+1},b_k)$ may be empty, as $\,a_k=a_{k+1}\,$ for some $k\geq1$ implies that $\,b_k=b_{k+1}$ and vice versa due to relation (\ref{b-r-f}) valid for all $k\in I_{n-1}$ (including $k=1$).\footnote{Note that $b_1$ in Lemma \ref{ml2} and $b_1$ in Lemma \ref{ml} are different (in contrast to $b_2$,$b_3$,...).}
\end{remark}\smallskip

\emph{Proof.} If $\,n<+\infty$ then the claim of the lemma can be proved by using Kuhn–Tucker theorem. We will give a direct general proof applicable to both cases
$n<+\infty$ and $n=+\infty$. The particular difficulty of the (most important) case $n=+\infty$ consists in the necessity to analyse extremal values of the function
$\{x_i\}_{i=1}^{+\infty}\mapsto\sum_{i=1}^{+\infty}e^{-bx_i}$ on the set of \emph{unbounded} nondecreasing sequences $\{x_i\}_{i=1}^{+\infty}$ of nonnegative numbers such that $\sum_{i=1}^{+\infty}a_ix_i=1$ and $x_1\in[0,c]$.\smallskip

Assume first that the $n$-tuple  $\{a_i\}_{i=1}^{n}$ consists of different numbers ($a_i\neq a_j$ for all $i\neq j$). In this case
(\ref{b-r-f}) shows that $b_{k+1}<b_{k}$ for any $k\in I_{n-1}$.
\smallskip

Denote the \emph{convex} function $\,\bar{x}=\{x_i\}_{i=1}^{n}\mapsto\sum_{i=1}^{n}e^{-bx_i}\,$ by $f$. Introduce the \emph{convex} sets
$$
 \displaystyle \Pi\doteq\left\{\{x_i\}_{i=1}^{n}\,\left|\, x_i\in \mathbb{R},\; \sum_{i=1}^{n}a_ix_i=1\right.\right\}
$$
and
$$
 \displaystyle \Lambda_k\doteq\left\{\{x_i\}_{i=1}^{n}\,\left|\, x_i\in \mathbb{R},\;  x_1\geq0,...,x_k\geq0 \right.\right\},\quad k\in I_n.
$$
It is clear that $\,\Omega_{\rm c}\subsetneq\Pi\cap\Lambda_k$ for any $k\in I_n$.

\smallskip
We prove the main claim of the lemma in $\,n+1\,$ steps (Steps 1,2,3.. and Step C), by showing its validity for $b\in[b_{k},b_{k-1})$ on Step $k$ and considering the case $b\in[b_0,+\infty)$, $b_0<+\infty$, on Step C.
\smallskip

\emph{Step 1.} Lemma \ref{ml-2} below with $k=0$ implies that $\,\inf\{f(\bar{x})\,|\,\bar{x}\in \Pi\}=f(\bar{x}_0)$, where
$\bar{x}_0=\{x^0_i\}_{i=1}^{n}$ is the $n$-tuple in $\Pi$ with the entries
\begin{equation}\label{2x0}
x^0_i=\frac{1}{b}\left( \frac{b-s_0}{d_0} -\ln a_i\right)\quad i\in I_n.
\end{equation}
Indeed, it is easy to see that $\,e^{-bx^0_i}=e^{\frac{s_0-b}{d_0}}\, a_i$ for any $i\in I_n$.
If $b\geq b_1$ then $x^0_i\geq0$ for any $i\in I_n$ and, hence, $\bar{x}_0$ is the minimum point
of the function $f$ on $\Pi\cap\Lambda_n$. If, in addition, $b\leq b_0$ then $x^0_1\leq c$. So,
$\bar{x}_0\in\Omega_{\rm c}$ is the minimum point of the function $f$ on $\Omega_{\rm c}\subset\Pi\cap\Lambda_n$. Hence, $\bar{x}_b=\bar{x}_0$ for $b\in [b_1,b_0]$.\smallskip

\emph{Step 2.} Assume that $b<b_1$. Then $x^0_1<0$. We will show that
\begin{equation}\label{st2-1}
\inf\{f(\bar{x})\,|\,\bar{x}\in \Pi\cap\Lambda_1\}=f(\bar{x}_1),
\end{equation}
where $\bar{x}_1=\{x^1_i\}_{i=1}^{n}$ is the $n$-tuple in $\Pi\cap\Lambda_1$ with the entries
$$
x^1_1=0,\quad x^1_i=\frac{1}{b}\left( \frac{b-s_1}{d_1} -\ln a_i\right)\quad i\in I_n\cap[2,+\infty).
$$
Assume that $f(\bar{y})<f(\bar{x}_1)$ for some $\bar{y}\in\Pi\cap\Lambda_1$. Since $x^0_1<0$ and $y_1\geq0$, there is $t\in[0,1]$ such that
$tx^0_1+(1-t)y_1=0$. Let $\bar{y}_1=t\bar{x}_0+(1-t)\bar{y}=\{0,y^1_2,y^1_3,...\}\in\Pi\cap\Lambda_1$. By the convexity of $f$ we have
\begin{equation}\label{st2-2}
f(\bar{y}_1)\leq tf(\bar{x}_0)+(1-t)f(\bar{y})\leq f(\bar{y}),
\end{equation}
where the last inequality holds because $\bar{x}_0$ is the minimum point of $f$ on $\Pi$ (by Step 1) and, hence, $f(\bar{x}_0)\leq f(\bar{y})$.

It is easy to see that $\,e^{-bx^1_i}=e^{\frac{s_1-b}{d_1}}\, a_i$ for any $i\in I_n\cap[2,+\infty)$. So, since $y_1^1=0$,
Lemma \ref{ml-2} below with $k=1$ implies that $f(\bar{y}_1)\geq f(\bar{x}_1)$. This and (\ref{st2-2}) contradict  the assumption $f(\bar{y})<f(\bar{x}_1)$.
Thus, (\ref{st2-1}) is proved. If $b\in[b_2,b_1)$ then $x^1_1=0\leq c$ and $x^1_i\geq0$ for any $i\in I_n$ and, hence, $\bar{x}_1\in\Omega_{\rm c}$ is the minimum point
of the function $f$ on $\Omega_{\rm c}\subset\Pi\cap\Lambda_n$. So, $\bar{x}_b=\bar{x}_1$ for $b\in[b_2,b_1)$.\smallskip

\emph{Step 3.}\footnote{Formally, Step 3 is not necessary: one can directly pass to the induction process considered below. We describe this step to show
all the basic tricks used in the induction process.} Assume that $b< b_2$. Then $x^1_2<0$. We will show that
\begin{equation}\label{st3-1}
\inf\{f(\bar{x})\,|\,\bar{x}\in \Pi\cap\Lambda_2\}=f(\bar{x}_2),
\end{equation}
where $\bar{x}_2=\{x^2_i\}_{i=1}^{n}$ is the $n$-tuple in $\Pi\cap\Lambda_2$ with the entries
$$
x^2_1=x^2_2=0,\quad x^2_i=\frac{1}{b}\left( \frac{b-s_2}{d_2} -\ln a_i\right)\quad i\in I_n\cap[3,+\infty).
$$
Assume that $f(\bar{y})<f(\bar{x}_2)$ for some $\bar{y}=\{y_1,y_2,...\}\in\Pi\cap\Lambda_2$. Since $x^1_2<0$ and $y_2\geq0$, there is $t\in[0,1]$ such that
$tx^1_2+(1-t)y_2=0$. Let $\bar{y}_1=t\bar{x}_1+(1-t)\bar{y}=\{y_1,0,y^1_3,...\}\in\Pi\cap\Lambda_2$. By the convexity of $f$ we have
\begin{equation}\label{st3-2}
f(\bar{y}_1)\leq tf(\bar{x}_1)+(1-t)f(\bar{y})\leq f(\bar{y}),
\end{equation}
where the last inequality holds because $\bar{x}_1$ is the minimum point of $f$ on $\Pi\cap\Lambda_1$ (by Step 2)  and, hence, $f(\bar{x}_1)\leq f(\bar{y})$.\smallskip

Let $\bar{y}_2=\{0,\frac{a_1y_1}{a_2},y^1_3,...\}\in\Pi\cap\Lambda_2$. Then $f(\bar{y}_2)<f(\bar{y}_1)$, since $a_1>a_2$ by the assumption.
By applying to $\bar{y}_2$ the trick previously applied to $\bar{y}$, we construct $\bar{y}_3=t'\bar{x}_1+(1-t')\bar{y}_2=\{0,0,y^3_3,...\}\in\Pi\cap\Lambda_2$ such that
$f(\bar{y}_3)\leq f(\bar{y}_2)$.

It is easy to see that $\,e^{-bx^2_i}=e^{\frac{s_2-b}{d_2}}\, a_i\,$ for any $\,i\in I_n\cap[3,+\infty)$. So, since $y^3_1=y^3_2=0$,
Lemma \ref{ml-2} below with $k=2$ implies that $f(\bar{y}_3)\geq f(\bar{x}_2)$. This, (\ref{st3-2}) and the inequalities after it contradict the assumption $f(\bar{y})<f(\bar{x}_2)$.
Thus, (\ref{st3-1}) is proved. If $b\in[b_3,b_2)$ then $\,x^2_1=0\leq c\,$ and $\,x^2_i\geq0\,$ for any $i\in I_n$ and, hence, $\bar{x}_2\in\Omega_{\rm c}$ is the minimum point
of the function $f$ on $\Omega_{\rm c}\subset\Pi\cap\Lambda_n$. So, $\bar{x}_b=\bar{x}_2$ for $b\in[b_3,b_2)$.\smallskip

Then we proceed by induction. Assume that at Step $k<n$ we have proved that
\begin{equation}\label{stk}
\inf\{f(\bar{x})\,|\,\bar{x}\in \Pi\cap\Lambda_{k-1}\}=f(\bar{x}_{k-1})
\end{equation}
for any $b<b_{k-1}$, where $\bar{x}_{k-1}=\{x^{k-1}_i\}_{i=1}^{n}$ is the $n$-tuple in $\Pi\cap\Lambda_{k-1}$ with the entries
\begin{equation*}
x^{k-1}_1=..=x^{k-1}_{k-1}=0,\quad x^{k-1}_i=\frac{1}{b}\left( \frac{b-s_{k-1}}{d_{k-1}} -\ln a_i\right)\quad i\in I_n\cap[k,+\infty).
\end{equation*}

\emph{Step $k+1$.} Assume that $b<b_k$. Then $x^{k-1}_k<0$. We will show that
\begin{equation}\label{stk+1}
\inf\{f(\bar{x})\,|\,\bar{x}\in \Pi\cap\Lambda_{k}\}=f(\bar{x}_k),
\end{equation}
where $\bar{x}_k=\{x^k_i\}_{i=1}^{n}$ is the $n$-tuple in $\Pi\cap\Lambda_k$ with the entries
\begin{equation}\label{xk}
x^k_1=..=x^k_k=0,\quad x^k_i=\frac{1}{b}\left( \frac{b-s_k}{d_k} -\ln a_i\right)\quad i\in I_n\cap[k+1,+\infty).
\end{equation}
Assume that $f(\bar{y})<f(\bar{x}_k)$ for some $\bar{y}=\{y_1,y_2,...\}\in\Pi\cap\Lambda_k$. Since $x^{k-1}_k<0$ and $y_k\geq0$, there is $t\in[0,1]$ such that
$tx^{k-1}_k+(1-t)y_k=0$. Let $\bar{y}_1=t\bar{x}_{k-1}+(1-t)\bar{y}=\{y_1,..,y_{k-1},0,y^1_{k+1},..\}\in\Pi\cap\Lambda_k$. By the convexity of $f$ we have
\begin{equation}\label{stk+2}
f(\bar{y}_1)\leq tf(\bar{x}_{k-1})+(1-t)f(\bar{y})\leq f(\bar{y}),
\end{equation}
where the last inequality holds because $\bar{x}_{k-1}$ is the minimum point of $f$ on $\Pi\cap\Lambda_{k-1}$ (by Step $k$)   and, hence, $f(\bar{x}_{k-1})\leq f(\bar{y})$.
\smallskip

Let $\bar{y}_2=\{0,y_2,..,y_{k-1},\frac{a_1y_1}{a_{k}},y^1_{k+1},...\}\in\Pi\cap\Lambda_k$. Then $f(\bar{y}_2)<f(\bar{y}_1)$, since $a_1>a_{k}$ by the assumption.
By applying to $\bar{y}_2$ the trick previously applied to $\bar{y}$, we construct the $n$-tuple $\bar{y}_3=t'\bar{x}_{k-1}+(1-t')\bar{y}_2=\{0,y_2,..,y_{k-1},0,y^3_{k+1},y^3_{k+2},...\}\in\Pi\cap\Lambda_k$ such that
$f(\bar{y}_3)\leq f(\bar{y}_2)$.

Acting according to this scheme, we construct the $n$-tuples
$$
\begin{array}{c}
  \bar{y}_4=\{0,0,y_3,..,y_{k-1},\frac{a_2y_2}{a_k},y^3_{k+1},y^3_{k+2},...\},\quad \bar{y}_5=\{0,0,y_3,..,y_{k-1},0,y^5_{k+1},y^5_{k+2},...\},
\\.....................................\\
\bar{y}_{2t}=\{0,..,0,y_{t+1},..,y_{k-1},\frac{a_ty_t}{a_k},y^{2t-1}_{k+1},...\},\quad \bar{y}_{2t+1}=\{0,..,0,y_{t+1},..,y_{k-1},0,y^{2t+1}_{k+1},...\},
\\\\t=3,4,..,k-2,\\\\
\bar{y}_{2k-2}=\{0,..,0,\frac{a_{k-1}y_{k-1}}{a_k},y^{2k-3}_{k+1},y^{2k-3}_{k+2},...\},\quad \bar{y}_{2k-1}=\{0,..,0,y^{2k-1}_{k+1},y^{2k-1}_{k+2},...\}
\end{array}
$$
in $\Pi\cap\Lambda_k$ such that
$$
f(\bar{y}_3)\geq f(\bar{y}_4)\geq f(\bar{y}_5)\geq..\geq f(\bar{y}_{2t})\geq f(\bar{y}_{2t+1})\geq..\geq f(\bar{y}_{2k-2})\geq f(\bar{y}_{2k-1}).
$$

It is easy to see that $\,e^{-bx^k_i}=e^\frac{s_k-b}{d_k}\, a_i$ for any $i\in I_n\cap[k+1,+\infty)$. So, since $y^{2k-1}_1=..=y^{2k-1}_k=0$,
Lemma \ref{ml-2} below implies that $f(\bar{y}_{2k-1})\geq f(\bar{x}_k)$. This, (\ref{stk+2}) and the inequalities after it contradict the assumption $f(\bar{y})<f(\bar{x}_k)$.
Thus, (\ref{stk+1}) is proved. If $b\in[b_{k+1},b_k)$ then $x^k_1=0\leq c$ and $x^k_i\geq0$ for any $i\in I_n$ and, hence, $\bar{x}_k\in\Omega_{\rm c}$ is the minimum point
of the function $f$ on $\Omega_{\rm c}\subset\Pi\cap\Lambda_n$. So, $\bar{x}_b=\bar{x}_k$ for $b\in[b_{k+1},b_k)$. This completes the induction process and the proof of the main claim of the lemma for any $b\leq b_0$ (in the case $a_i\neq a_j$ for all $i\neq j$).\smallskip

\emph{Step C.} Assume that $b_0<+\infty$ and  $b>b_0$. Then $x^{0}_1>c$. We will show that
\begin{equation}\label{stk-c}
\inf\{f(\bar{x})\,|\,\bar{x}\in \Omega_{\rm c}\}=f(\bar{x}_c),
\end{equation}
where $\bar{x}_c=\{x^c_i\}_{i=1}^{n}$ is the $n$-tuple in $\Omega_{\rm c}$ with the entries
\begin{equation*}
x^c_1=c,\quad x^c_i=\frac{1}{b}\left( \frac{b(1-a_1c)-s_1}{d_1} -\ln a_i\right)\quad i\in I_n\cap[2,+\infty).
\end{equation*}
Assume that $f(\bar{y})<f(\bar{x}_c)$ for some $\bar{y}=\{y_1,y_2,...\}\in\Omega_{\rm c}$. Since $x^{0}_1>c$ and $y_1\in[0,c]$, there is $t\in[0,1]$ such that
$\,tx^{0}_1+(1-t)y_1=c$. Let $\bar{y}_1=t\bar{x}_{0}+(1-t)\bar{y}=\{c,y^1_{2},y^1_3,..\}\in\Omega_{\rm c}$. By the convexity of $f$ we have
\begin{equation}\label{stk+c}
f(\bar{y}_1)\leq tf(\bar{x}_{0})+(1-t)f(\bar{y})\leq f(\bar{y}),
\end{equation}
where the last inequality holds because $\bar{x}_{0}$ is the minimum point of $f$ on $\Pi\cap\Lambda_{n}$ (by Step $1$) and, hence, $\,f(\bar{x}_{0})\leq f(\bar{y})$.

It is easy to see that $\,e^{-bx^c_i}=e^{\frac{s_1-b(1-a_1c)}{d_1}}\, a_i$ for any $i\in I_n\cap[2,+\infty)$. So, since $y_{1}^1=x_{1}^c=c$,
Lemma \ref{ml-2} below with $k=1$ implies that $f(\bar{y}_{1})\geq f(\bar{x}_c)$. This and (\ref{stk+c})  contradict the assumption $f(\bar{y})<f(\bar{x}_c)$.
Thus, (\ref{stk-c}) is proved and hence $\bar{x}_b=\bar{x}_c$ for $b\in(b_{0},+\infty)$.
\smallskip

Assume now that the $n$-tuple  $\{a_i\}_{i=1}^{n}$ contains equal numbers (i.e. $a_i=a_{i+1}$ for some $i\in I_n$ which means that $b_i=b_{i+1}$ due to (\ref{b-r-f})). In this case we may repeat
the above steps with the appropriate modification used Lemma \ref{ml-3} below.
We will show how Steps $2$ and $k+1$ should be modified. Steps $1$ and $C$ remain the same.

Step $2$ should be modified if  $\,a_{1}=a_{2}=..=a_{m}>a_{m+1}$ for some  $m>1$  and, hence, (\ref{b-r-f}) shows that $b_{1}=b_{2}=..=b_{m}>b_{m+1}$. We may exclude the case $\,a_{1}=a_{2}=..=a_{n}$, $n<+\infty$,  since in this  case $\,b_{1}=b_{2}=..=b_{n}=0$. In Step $1$ we have proved that
\begin{equation}\label{2stk+}
\inf\{f(\bar{x})\,|\,\bar{x}\in \Pi\}=f(\bar{x}_{0}),
\end{equation}
where  $\bar{x}_{0}=\{x^{0}_i\}_{i=1}^{n}$ is the $n$-tuple in $\Pi$ defined in (\ref{2x0}).\smallskip

\emph{Modified step $2$.}
Let $b<b_{1}=b_{m}$. Then  $x^{0}_{1}=..=x^{0}_{m}<0$. We will show that
\begin{equation}\label{2stk+1+}
\inf\{f(\bar{x})\,|\,\bar{x}\in \Pi\cap\Lambda_{m}\}=f(\bar{x}_{m}),
\end{equation}
where $\bar{x}_{m}=\{x^{m}_i\}_{i=1}^{n}$ is the $n$-tuple in $\Pi\cap\Lambda_{m}$ defined in (\ref{xk}) with $k=m$.

Suppose that $f(\bar{y})<f(\bar{x}_{m})$ for some $\bar{y}=\{y_1,y_2,...\}\in\Pi\cap\Lambda_{m}$. By Lemma \ref{ml-3} below
we may assume that $y_{1}=...=y_{m}\geq0$. Since $x^{0}_1=..=x^{0}_{m}<0$, there is $t\in[0,1]$ such that
$tx^{0}_i+(1-t)y_i=0$, $i=\overline{1,m}$. Let $\bar{y}_1=t\bar{x}_{0}+(1-t)\bar{y}=\{0,..,0,y^1_{m+1},y^1_{m+2},..\}\in\Pi\cap\Lambda_{m}$. By the convexity of $f$ we have
\begin{equation}\label{2stk+2+}
f(\bar{y}_1)\leq tf(\bar{x}_{0})+(1-t)f(\bar{y})\leq f(\bar{y}),
\end{equation}
where the last inequality holds because $\bar{x}_{0}$ is the minimum point of $f$ on $\Pi$ (by Step $1$) and, hence, $f(\bar{x}_{0})\leq f(\bar{y})$.

It is easy to see that $\,e^{-bx^m_i}=e^{\frac{s_m-b}{d_m}}\, a_i$ for any $i\in I_n\cap[m+1,+\infty)$. So, since $y_{1}^1=..=y_{m}^1=0$,
Lemma \ref{ml-2} below with $k=m$ implies that $f(\bar{y}_{1})\geq f(\bar{x}_m)$. This and (\ref{2stk+2+}) contradict the assumption $f(\bar{y})<f(\bar{x}_m)$.
Thus, (\ref{2stk+1+}) is proved. If $b\in[b_{m+1},b_1=b_{m})$ then $x^m_1=0\leq c$ and $x^m_i\geq0$ for any $i\in I_n$ and, hence, $\bar{x}_m\in\Omega_{\rm c}$ is the minimum point
of the function $f$ on $\Omega_{\rm c}\subset\Pi\cap\Lambda_n$. So, $\bar{x}_b=\bar{x}_m$ for $b\in[b_{m+1},b_1=b_{m})$.

\smallskip

To show how Step $k+1$ should be modified assume that $a_{j-1}>a_{j}=a_{j+1}=..=a_{j+m}>a_{j+m+1}$ for some $j>1$ and $m>0$. Hence, (\ref{b-r-f}) shows that $b_{j-1}>b_{j}=b_{j+1}=..=b_{j+m}>b_{j+m+1}$. We may exclude the case $j+m=n<+\infty$,  since in this  case $b_{j}=b_{j+1}=..=b_{j+m}=0$.
Assume that in Step $k\leq j$ we have proved that
\begin{equation}\label{stk+}
\inf\{f(\bar{x})\,|\,\bar{x}\in \Pi\cap\Lambda_{j-1}\}=f(\bar{x}_{j-1}),
\end{equation}
for any $b<b_{j-1}$, where  $\bar{x}_{j-1}=\{x^{j-1}_i\}_{i=1}^{n}$ is the $n$-tuple in $\Pi\cap\Lambda_{j-1}$ defined in (\ref{xk}) with $k=j-1$.\footnote{$k=j$ if  either $j=1$ or $a_i\neq a_l$ for all $i>l>j$ (since this means that $b_{j-1}<b_{j-2}<...<b_1$).} \smallskip

\emph{Modified step $k+1$.}
Let $b<b_{j}=b_{j+m}$. Then  $x^{j-1}_{j}=..=x^{j-1}_{j+m}<0$. We will show that
\begin{equation}\label{stk+1+}
\inf\{f(\bar{x})\,|\,\bar{x}\in \Pi\cap\Lambda_{j+m}\}=f(\bar{x}_{j+m}),
\end{equation}
where $\bar{x}_{j+m}=\{x^{j+m}_i\}_{i=1}^{n}$ is the $n$-tuple in $\Pi\cap\Lambda_{j+m}$ defined in (\ref{xk}) with $k=j+m$.

Suppose that $f(\bar{y})<f(\bar{x}_{j+m})$ for some $\bar{y}=\{y_1,y_2,...\}\in\Pi\cap\Lambda_{j+m}$. By Lemma \ref{ml-3} below
we may assume that $y_{j}=...=y_{j+m}\geq0$. Since $x^{j-1}_j=..=x^{j-1}_{j+m}<0$, there is $t\in[0,1]$ such that
$tx^{j-1}_i+(1-t)y_i=0$, $i=\overline{j,j+m}$. Let $\bar{y}_1=t\bar{x}_{j-1}+(1-t)\bar{y}=\{y_1,..,y_{j-1},0,..,0,y^1_{j+m+1},y^1_{j+m+2},...\}\in\Pi\cap\Lambda_{j+m}$. By the convexity of $f$ we have
\begin{equation}\label{stk+2+}
f(\bar{y}_1)\leq tf(\bar{x}_{j-1})+(1-t)f(\bar{y})\leq f(\bar{y}),
\end{equation}
where the last inequality holds because $\bar{x}_{j-1}$ is the minimum point of $f$ on $\Pi\cap\Lambda_{j-1}$ (by Step $k$) and, hence, $f(\bar{x}_{j-1})\leq f(\bar{y})$.\smallskip

Let $\bar{y}_2=\{0,y_2,..,y_{j-1},\frac{a_1y_1}{(m+1)a_{j}},..,\frac{a_1y_1}{(m+1)a_{j}}, y^1_{j+m+1},y^1_{j+m+2},...\}\in\Pi\cap\Lambda_{j+m}$. Then Lemma \ref{ml-3} below shows that $f(\bar{y}_2)<f(\bar{y}_1)$, since $a_1>a_{j}$ by the assumption. By applying to $\bar{y}_2$ the trick previously applied to $\bar{y}$, we construct the $n$-tuple $\bar{y}_3=t'\bar{x}_{j-1}+(1-t')\bar{y}_2=\{0,y_2,..,y_{j-1},0,..,0,y^3_{j+m+1},y^3_{j+m+2},...\}\in\Pi\cap\Lambda_{j+m}$ such that
$f(\bar{y}_3)\leq f(\bar{y}_2)$.

Acting according to this scheme, we construct the $n$-tuples
$$
\begin{array}{c}
  \bar{y}_4=\left\{0,0,y_3,..,y_{j-1},\frac{a_2y_2}{(m+1)a_{j}},..,\frac{a_2y_2}{(m+1)a_{j}},y^3_{j+m+1},y^3_{j+m+2},...\right\},\\\\ \bar{y}_5=\{0,0,y_3,..,y_{j-1},0,..,0,y^5_{j+m+1},y^5_{j+m+2},...\},
\\.....................................\\
\bar{y}_{2t}=\left\{0,..,0,y_{t+1},..,y_{j-1},\frac{a_ty_t}{(m+1)a_{j}},..,\frac{a_ty_t}{(m+1)a_{j}},y^{2t-1}_{j+m+1},y^{2t-1}_{j+m+2},...\right\},\\\\ \bar{y}_{2t+1}=\{0,..,0,y_{t+1},..,y_{j-1},0,..,0,y^{2t+1}_{j+m+1},y^{2t+1}_{j+m+2},...\},
\\\\t=3,4,..,j-2,\\\\
\bar{y}_{2j-2}=\left\{0,..,0,\frac{a_{j-1}y_{j-1}}{(m+1)a_j},..,\frac{a_{j-1}y_{j-1}}{(m+1)a_j},y^{2j-3}_{j+m+1},y^{2j-3}_{j+m+2},...\right\},\\\\
\bar{y}_{2j-1}=\{0,..,0,y^{2j-1}_{j+m+1},y^{2j-1}_{j+m+2},...\}
\end{array}
$$
in $\Pi\cap\Lambda_{j+m}$ such that
$$
f(\bar{y}_3)\geq f(\bar{y}_4)\geq f(\bar{y}_5)\geq..\geq f(\bar{y}_{2t})\geq f(\bar{y}_{2t+1})\geq..\geq f(\bar{y}_{2j-2})\geq f(\bar{y}_{2j-1}).
$$

It is easy to see that $\,e^{-bx^{j+m}_i}=e^{\frac{s_{j+m}-b}{d_{j+m}}}\, a_i$ for any $i\in I_n\cap[j+m+1,+\infty)$. So, since $y^{2j-1}_1=..=y^{2j-1}_{j+m}=0$,
Lemma \ref{ml-2} below with $k=j+m$ implies that $f(\bar{y}_{2j-1})\geq f(\bar{x}_{j+m})$. This, (\ref{stk+2+}) and the inequalities after it contradict the assumption $f(\bar{y})<f(\bar{x}_{j+m})$.
Thus, (\ref{stk+1+}) is proved. If $b\in[b_{j+m+1},b_j=b_{j+m})$ then $x^{j+m}_1=0\leq c$ and $x^{j+m}_i\geq0$ for any $i\in I_n$ and, hence, $\bar{x}_{j+m}\in\Omega_{\rm c}$ is the minimum point
of the function $f$ on $\Omega_{\rm c}\subset\Pi\cap\Lambda_n$. So, $\bar{x}_b=\bar{x}_{j+m}$ for $b\in[b_{j+m+1},b_j=b_{j+m})$. \smallskip

To complete the proof it suffices to note that
\begin{itemize}
  \item  the uniqueness if the point at which the infimum in  the r.h.s. of (\ref{ml+}) is attained
follows from the strict convexity of the function $e^x$ on $\mathbb{R}$;
  \item the continuity of the function $\,b\mapsto x^b_i\,$ on $\,(0,+\infty)\,$ for each $\,i\in I_n$ is proved by showing
  its continuity at the "gluing points" $b_0,b_1,b_2,...$.
\end{itemize}
$\Box$ \medskip

\begin{lemma}\label{ml-2} \emph{Assume that the notation and conditions of Lemma \ref{ml} are valid. Let $\,\bar{x}=\{x_i\}_{i=1}^n$ be an $n$-tuple of nonnegative  numbers such that $\,\sum_{i=1}^n a_ix_i=1$
and $\,e^{-bx_i}=\lambda a_i$ for all $\,i\in I_n\cap(k,+\infty)$ and some $\lambda>0$, where $k\in\{0\}\cup I_n$. Then
$$
f(\bar{x})\leq f(\bar{y})
$$
for any $n$-tuple $\,\bar{y}=\{y_i\}_{i=1}^n$ of nonnegative numbers such that $\sum_{i=1}^n a_iy_i=1$  and $y_i=x_i$ for all $i\in I_n\cap[1,k]$ (if $\,k=0\,$ then the last condition holds trivially).}
\end{lemma}\smallskip

\emph{Proof.} Let $t_i=y_i-x_i$ for all $i\in I_n$. Then  $\sum_{i=k+1}^na_it_i=\sum_{i=k+1}^na_iy_i-\sum_{i=k+1}^na_ix_i=0$, and hence
$$
f(\bar{y})-f(\bar{x})=\sum_{i=k+1}^ne^{-bx_i}(e^{-bt_i}-1)\geq \sum_{i=k+1}^ne^{-bx_i}(-bt_i)=-\lambda b\sum_{i=k+1}^na_it_i=0,
$$
where the inequality $e^z\geq 1+z$ (valid for any $z\in \mathbb{R}$)  is used. In the case $n=+\infty$ we have to note that all the above series are absolutely convergent.  $\Box$\smallskip

\begin{lemma}\label{ml-3} \emph{Let $\{a_i\}_{i=1}^{n}$ be an $n$-tuple of positive numbers arranged in the non-increasing order such that
$\,a_k=a_{k+1}=...=a_{k+m}$ for some $k$ and $m$. If $\,\bar{x}=\{x_i\}_{i=1}^n$ is an $n$-tuple of nonnegative  numbers such that $\,\sum_{i=1}^n a_ix_i=1$
then
$$
\sum_{i=1}^n a_iy_i=1\quad and \quad \sum_{i=1}^{n}e^{-bx_i}\geq \sum_{i=1}^{n}e^{-by_i},
$$
where}
$$
y_i=\left\{\begin{array}{ll}
        \frac{1}{m+1}\sum_{i=k}^{k+m}x_i &\textrm{if}\;\;  k\leq i\leq k+m \\
        x_i & \textrm{otherwise}.
       \end{array}\right.
$$
\end{lemma}\smallskip

\emph{Proof.} It suffices to note that $e^x$ is a convex function. $\Box$ \medskip

\subsection*{A-2. Proof of Lemma \ref{V} for $\mathfrak{H}$-operators of type $A$}

Let $H$ be an $\mathfrak{H}$-operator o type $A$ with representation (\ref{H-form})
in which $\,h_n<+\infty\,$ but $\,h_k=+\infty\,$ for all $k>n$. In fact, we may assume that
$H$ is a positive operator on the finite-dimensional space $\,\H_n\doteq\D(H)\,$ with the
representation
$$
H=\sum_{k=1}^{n} h_k |\tau_k\rangle\langle\tau_k|,
$$
where $\left\{\tau_k\right\}_{k=1}^{n}$ is the basis in $\H_n$ consisting of eigenvectors of $H$. Let $E>h_1$ be arbitrary.

\smallskip

If $E< h_*(H)\doteq\frac{1}{n}\sum_{k=1}^nh_k$ then simple arguments (cf. the proof of Proposition 1 in \cite{EC}) show that
the equation
\begin{equation}\label{mlk}
\Tr H e^{-\beta H}=E \Tr e^{-\beta H}
\end{equation}
has a unique positive solution $\beta_H(E)$. Let
\begin{equation*}
\gamma_H(E)=\frac{e^{-\beta_H(E) H}}{\Tr e^{-\beta_H(E)H}}\in\S(\H_n).
\end{equation*}
Then for any state $\rho$ in $\S(\H_n)$ with $\Tr H\rho=E$ we have
$$
S(\rho)=E\beta_H(E)+\ln \Tr e^{-\beta_H(E)H}-
D(\rho\shs\|\gamma_H(E))\leq E\beta_H(E)+\ln \Tr e^{-\beta_H(E)H}=S(\gamma_H(E))
$$
by nonnegativity of the quantum relative entropy denoted by $D(\cdot\|\cdot)$. This shows that
$$
F_H(E)=S(\gamma_H(E))
$$
because it is easy to see that the supremum in the definition (\ref{F-def}) can be taken only
over all states $\rho$ with $\Tr H\rho=E$.

The function $G(\beta)=E\beta+\ln \Tr e^{-\beta H}$ is strictly convex on $[0,+\infty)$. Indeed, it is easy to see that
$$
G''(\beta)=\frac{\Tr H^2e^{-\beta H}}{\Tr e^{-\beta H}}-\left[\frac{\Tr He^{-\beta H}}{\Tr e^{-\beta H}}\right]^2>0,
$$
since the r.h.s. of this expression can be represented as $\mathbb{E}(\xi^2)-[\mathbb{E}(\xi)]^2$, where
$\xi$ is a random variable taking the values $h_1,...,h_n$ with the probabilities $e^{-\beta h_1}/c,...,e^{-\beta h_n}/c$, $c=\sum_{k=1}^{n}e^{-\beta h_k}$.\smallskip

Note also that
$$
G'(\beta)=E-\frac{\Tr He^{-\beta H}}{\Tr e^{-\beta H}}.
$$
So, if $E<h_*(H)$ then $G(0)=E-h_*(H)<0$ and, hence, the convex continuously differentiable function $G$ takes its minimal value on $[0,+\infty)$ at the point $\beta=\beta_H(E)$, since $G'(\beta_H(E))=0$. Thus,
$$
\min_{\beta\geq 0}G(\beta)=G(\beta_H(E))=E\beta_H(E)+\ln \Tr e^{-\beta_H(E)H}=F_H(E).
$$
If $E\geq h_*(H)$ then $\Tr H\bar{\rho}_n\leq E$, where $\bar{\rho}_n=\frac{1}{n}\sum_{k=1}^{n}|\tau_k\rangle\langle\tau_k|$ is the chaotic state
in $\S(\H_n)$. Hence,
$$
F_H(E)=\ln n=G(0)=\min_{\beta\geq 0}G(\beta),
$$
where the last equality holds because in this case $\,G'(0)=E-h_*(H)\geq0\,$ and due to the convexity of $G$.

Thus, the expression
$$
F_H(E)=\inf_{\beta\geq 0}\left( E\beta +\ln\Tr e^{-\beta H}\right)
$$
is proved for any $E>h_1$. For $E=h_1$ this expression holds trivially, since it easy to see that $\,F_H(h_1)=\ln\dim\ker(H-h_1I_{\H_n})$.

\bigskip\bigskip

I am grateful to A.S.Holevo and to the participants of his seminar  "Quantum Probability, Statistics, Information" (the Steklov  Mathematical Institute) for useful discussion.


\begin{thebibliography}{99}



\bibitem{D++} Audenaert,~K.M.R., Bergh,~B., Datta,~N., Jabbour,~M.G., Capel,~Á., Gondolf,~P., "Continuity bounds for quantum entropies arising from a fundamental entropic inequality", arXiv:2408.15306.

\bibitem{D+++} Berta,~M., Lami,~L., Tomamichel,~M., "Continuity of entropies via integral representations", IEEE Trans. Inf. Theory 71(3), 1896-1908 (2025); arXiv:2408.15226.

\bibitem{BDJ} Becker,~S., Datta,~N., Jabbour,~M.G.. "From Classical to Quantum: Uniform Continuity Bounds on Entropies in Infinite Dimensions",
IEEE Transactions on Information Theory, \textbf{69}(7), pp. 4128-4144 (2023); arXiv:2104.02019 v.2

\bibitem{BDJSH} Becker,~S., Datta,~N., Jabbour,~M.G.,  Shirokov,~M.E. , "Optimal continuity bound for the von Neumann
 entropy under energy constraints",  arXiv:2410.02686v2.

\bibitem{Bennett} Bennett,~C.H., DiVincenzo,~D.P., Smolin,~J.A., Wootters,~W.K.: "Mixed-state entanglement and quantum error correction",  Phys. Rev. A \textbf{54}, 3824-3851 (1996).




\bibitem{H-SCI} Holevo,~A.S.: "Quantum systems, channels, information.
A mathematical introduction", Berlin, DeGruyter (2012).

\bibitem{4H} Horodecki R., Horodecki P., Horodecki M., Horodecki K. "Quantum entanglement", Rev. Mod. Phys.
2009. V.81, P.865-942.


\bibitem{Wilde-new} Khatri~S, Wilde~M~M, "Principles of Quantum Communication Theory: A Modern Approach", arXiv:2011.04672[quant-ph].

\bibitem{Mirsky} Mirsky,~L.: "Symmetric gauge functions and unitarily invariant norms", Quart. J. Math.Oxford \textbf{2}(11), 50-59 (1960).

\bibitem{Mirsky-rr} Ghourchian,~H., Gohari,~A., Amini,~A., "Existence and continuity of differential entropy for
a class of distributions", IEEE Commun. Lett.  \textbf{21}(7), 1469–1472 (2017).  https://doi.org/10.1109/LCOMM.2017.2689770.


\bibitem{Moser} Moser~S.M., "Information Theory (Lecture Notes)," 2018. [Online]. Available: https://moser-isi.ethz.ch/
docs/it script v616.pdf.


\bibitem{L-2} Lindblad,~G.: "Expectation and Entropy Inequalities for Finite
Quantum Systems", Commun. Math. Phys. 39(2), 111-119 (1974).


\bibitem{N&Ch} Nielsen,~M.A., Chuang,~I.L.: "Quantum Computation and Quantum
Information",  Cambridge University Press (2000).




\bibitem{O&P} Ohya,~M., Petz,~D.: "Quantum Entropy and Its Use", Theoretical and Mathematical Physics, Springer Berlin Heidelberg (2004).

\bibitem{P&V} Plenio,~M.B., Virmani,~S.: "An introduction to entanglement measures", Quantum Inf. Comput. \textbf{7}(1-2), 1-51 (2007).

\bibitem{EC} Shirokov,~M.E.: "Entropy characteristics of subsets of states I. Izv. Math. \textbf{70}(6), 1265-1292  (2006); arXiv: quant-ph/0510073.


\bibitem{QC} Shirokov,~M.E.: "Quantifying continuity of characteristics of composite quantum systems",  Phys. Scr. \textbf{98}, 042002 (2023).


\bibitem{LCB}  Shirokov,~M.E.: "Close-to-optimal continuity bound for the von Neumann entropy and other quasi-classical applications of the Alicki–Fannes–Winter technique", Lett. Math. Phys., \textbf{113},  121 , 35 pp. (2023); arXiv: 2207.08791.


\bibitem{LCB+}  Shirokov,~M.E.: "The Alicki-Fannes-Winter technique in the quasi-classical settings: advanced version and its applications",
Lobachevskii Journal of Mathematics, \textbf{46}(6), 2632–2658 (2025); arXiv:2505.00882.

\bibitem{ISB}  Shirokov,~M.E.: "On state-dependant improvements of the semicontinuity bounds for basic characteristics of quantum systems", to appear in the arXiv.

\bibitem{FCB} Shirokov,~M.E.: "Semicontinuity bounds for the von Neumann entropy and partial majorization", arXiv:2504.08098.




\bibitem{W} Wehrl,~A., "General properties of entropy", Rev. Mod. Phys. \textbf{50}, 221-250 (1978).

\bibitem{Wilde} Wilde,~M.M., "Quantum Information Theory", Cambridge, UK: Cambridge Univ. Press, (2013).

\bibitem{W-CB} Winter,~A.: "Tight uniform continuity bounds for quantum entropies: conditional entropy, relative entropy distance and energy constraints", Commun. Math. Phys. 347(1), 291-313 (2016); arXiv:1507.07775 (v.6).

\bibitem{Lami-new} Yamasaki,~H., Kuroiwa,~K., Hayden,~P., Lami,~L.: "Entanglement cost for infinite-dimensional physical systems",
arXiv:2401.09554 (2024).









\end{thebibliography}
\end{document}